\documentclass[english,aps,reprint]{revtex4}
\usepackage[T1]{fontenc}
\usepackage[latin9]{inputenc}
\usepackage{amsmath}
\usepackage{graphicx}
\usepackage{amssymb}
\usepackage{esint}

\makeatletter

\providecommand{\tabularnewline}{\\}

\@ifundefined{textcolor}{}
{%
 \definecolor{BLACK}{gray}{0}
 \definecolor{WHITE}{gray}{1}
 \definecolor{RED}{rgb}{1,0,0}
 \definecolor{GREEN}{rgb}{0,1,0}
 \definecolor{BLUE}{rgb}{0,0,1}
 \definecolor{CYAN}{cmyk}{1,0,0,0}
 \definecolor{MAGENTA}{cmyk}{0,1,0,0}
 \definecolor{YELLOW}{cmyk}{0,0,1,0}
 }

\@ifundefined{definecolor}
 {\usepackage{color}}{}
\@ifundefined{definecolor}
 {\@ifundefined{definecolor}
 {\usepackage{color}}{}
}{}
\makeatother

\makeatother

\usepackage{babel}

\makeatother

\usepackage{babel}

\begin{document}

\title{Finite energy spectral function of an anisotropic 2D system of coupled
Hubbard chains }

\author{P. Ribeiro}

\email{ribeiro@cfif.ist.utl.pt}

\affiliation{CFIF, Instituto Superior Técnico, TU Lisbon, Av. Rovisco Pais, 1049-001
Lisboa, Portugal}

\author{P. D. Sacramento}

\email{pdss@cfif.ist.utl.pt}

\affiliation{Departamento de F\'{i}sica and CFIF, Instituto Superior Técnico,
TU Lisbon, Av. Rovisco Pais, 1049-001 Lisboa, Portugal and\\
 Department of Physics, The Chinese University of Hong Kong, Hong
Kong, China}

\author{K. Penc}

\email{penc@szfki.hu }

\affiliation{Research Institute for Solid State Physics and Optics, P.O. Box 49,
H-1525 Budapest, Hungary }
\begin{abstract}
We study the crossover from the one-dimensional to the two-dimensional
Hubbard model in the photoemission spectra of weakly coupled chains.
The chains with on-site repulsion are treated using the spin-charge
factorized wave function, that is known to provide an essentially
exact description of the chain in the strong coupling limit. The hoppings
between the chains are considered as a perturbation. We calculate
the dynamical spectral function at all energies in the random-phase
approximation, by resuming an infinite set of diagrams. Even though
the hoppings drive the system from a fractionalized Luttinger-liquid-like
system to a Fermi-liquid-like system at low energies, significant
characteristics of the one-dimensional system remain in the two-dimensional
system. Furthermore, we find that introducing (frustrating) hoppings
beyond the nearest neighbor one, the interference effects increase
the energy and momentum range of the one--dimensional character. 
\end{abstract}
\maketitle
\global\long\def\ket#1{\left| #1\right\rangle }

\global\long\def\bra#1{\left\langle #1 \right|}

\global\long\def\kket#1{\left\Vert #1\right\rangle }

\global\long\def\bbra#1{\left\langle #1\right\Vert }

\global\long\def\braket#1#2{\left\langle #1\right. \left| #2 \right\rangle }

\global\long\def\bbrakket#1#2{\left\langle #1\right. \left\Vert #2\right\rangle }

\global\long\def\av#1{\left\langle #1 \right\rangle }

\global\long\def\tr{\text{Tr}}

\global\long\def\im{\text{Im}}

\global\long\def\re{\text{Re}}

\global\long\def\sign{\text{sign}}

\global\long\def\Det{\text{Det}}

\section{Introduction}

The Hubbard model is believed to contain most of the fundamental physics
of a great variety of materials ranging from weak interacting metals
to Mott insulators. It may also capture some of the phenomena responsible
for the high $T_{c}$ superconductivity of cuprates. In one spatial
dimension the Hubbard chain is exactly solvable by the Bethe ansatz
\cite{Lieb_1968}, furthermore, the low energy properties are understood
in details within the Luttinger Liquid (LL) theory \cite{Haldane}.
In particular, the elementary excitations turn out to be fractionalized,
carrying either charge or spin quantum numbers, a property that is
a rather generic feature in 1D strongly interacting electron systems
\cite{Book}. Thanks to the advent of bosonization the low energy
physics of 1D interacting electrons is now well understood including
the computation of many observables \cite{Book,Gogolin_1998,Voit,Frahm,CarmeloGS}.
However, despite the enormous success of these complementary approaches
the computation of observables for arbitrary energies was obtained
only in some restricted limits \cite{Penc_1995,Penc_1996,Halffilling}
or using some additional approximations \cite{Sing_2003,Carmelo_2004,Carmelo_2004_b,Carmelo_2006,Bozi_2008}.

In higher dimensions the physical picture is much less clear. It has
fueled controversy, mainly motivated by the hight Tc phenomena in
the layered cuprate oxides. Generically, in dimensions higher than
one, excitations are not fractionalized, the most well-known example
being the quasi-particles in a Fermi liquid (FL). A few remarkable
experimental and model examples exist, however, where the electrons
fractionalize. The most spectacular example is the fractional quantum
Hall effect, where fractionalization of quasiparticles has been predicted
theoretically\cite{FQHEthe} and consequently found experimentally
\cite{FQHEexp}. Fractionalization has also been theoretically proposed
in electron systems with frustrated nearest neighbor interactions
\cite{tVmodel}. Further examples include quantum spin-liquids, where
the presence of frustration may lead to deconfinement of the spinons
in the two-dimensional system\cite{Sachdev1}. For example, the quasi-two
dimensional triangular spin system Cs$_{2}$CuCl$_{4}$\cite{Coldea2d}
has an excitation spectrum that can be described, similarly to the
one dimensional case \cite{Heisenberg}, by a continuum originated
from fractionalized pairs of spin $1/2$ spinons\cite{Kohno_2007}.
This property has been verified experimentally for several quasi-one
dimensional spin $1/2$ systems like CPC \cite{CPC}, KCuF$_{3}$
\cite{Nagler,Tennant} and copper benzoate \cite{Dender}.

Experimentally, the single--particle properties of the material are
most directly measured by photoemission. The intensity of the extraction
of the electron by photon at given energy and momentum transfer is
directly proportional to the spectral function -- the imaginary part
of the one-particle Green's function. If most of the spectral weight
is carried by well defined spectral lines one expects excitations
to be sharp, electron-like coherent modes (quasiparticles). On the
contrary, broad continua signal fractionalization of the electronic
degrees of freedom.

Low energy descriptions for the two dimensional case have been proposed
that predict a fractionalization description of the low energy physics.
Experimentally such low energy features are difficult to observe in
the photo-emission data since they are obscured by resolution and
noise. Therefore, it is useful to have a prediction over the full
energy range to compare with experiments. The motivation for this
work is twofold: on the one hand to provide an approximate spectral
function valid to arbitrary energy and on the other to clarify the
role of frustration in the underlying excitation.

In this work we address the dimensional crossover from one to two
dimensions in a strongly correlated electron system by coupling Hubbard
chains within the random phase approximation (RPA). This approximation
leads to a description of the 2D quantities spectral function in terms
of the 1D Greens's function of the chain. Though coupled one-dimensional
chain tend to order at low temperature, in this work we will assume
that we are at sufficiently high energies $\omega$ and temperatures
$T$ , typically larger than some crossover temperature $T_{\text{1D}}$
after which LL perturbation should be valid. The general expression
obtained by the RPA for the two dimensional spectral function is valid
for any values of $U$ and filling factor, as well as for any kind
of small inter-chain hopping. In particular it is possible to study
the role of hopping in different geometries, highly frustrated cases
as well as non-frustrated ones. Due to the lack of theoretical expressions
for the spectral function in one dimensional for generic $U$ we concentrate
our study on the $U\to\infty$ limit using the results derived in
\cite{Penc_1997}. The exact results obtained for the spectral function
of the Hubbard chain in the $U\rightarrow\infty$ limit, can be extended
to finite but large $U$ and are used in the RPA to obtain the same
function in higher dimensions.

\vspace*{0.6cm}

Previous works have dealt with the issue of coupled LL or coupled
Hubbard chains. Contrasting with the LL-like features of decoupled
chains, FL behavior is generically expected for weakly interacting
systems and large inter-chain hopping terms. The interpolation between
LL and FL regimes as the inter-chain hopping increases, as well as
the energy scales for which each description is valid have been largely
discussed. 

Using perturbative renormalization group (RG) and an RPA-like expression
for the two-dimensional Green's function, it was shown \cite{Wen_1990},
starting from a LL, that the hopping is relevant if $\theta<1$ and
irrelevant if $\theta>1$, where $\theta$ is the LL exponent characterizing
the low frequency behavior of the density of states $N\left(\omega\right)\sim\left|\omega\right|^{\theta}$,
(note that $\theta=0$ corresponds to the non-interacting case). In
the first case the two-dimensional Green's function develops well
define poles near the Fermi energy with a nonzero quasiparticle residue
$\left(Z\right)$ for non-vanishing inter-chain hoppings and in the
second case $Z$ vanishes. In the same direction it was pointed out
that using a $d=1+\epsilon$ expansion that the only weak-coupling
fixed point for $\epsilon>0$ is the FL one \cite{Castellani_1994}.
Using a path integral formulation \cite{Boies_1995} (like RPA) the
results of \cite{Wen_1990} where rederived, but it was pointed out
that higher order processes could extend the FL behavior beyond $\theta=1$.
Subsequent works, using exact resummation of some infinite class of
diagrams \cite{Arrigoni_1998,Arrigoni_1999,Arrigoni_2000}, also corroborate
this result. It was also shown, using bosonization, that even if long
range 3D Coulomb interactions were considered the 1D LL regime leads
to a FL, for any hopping, but anomalous scaling was found in the FL
phase for small hoppings \cite{Kopietz_1995,Kopietz_1997}. 

The picture that FL behavior is obtained as soon as inter-chain hopping
is introduced has, however, to be interpreted as being valid only
above some finite energy scale. The introduction of inter-chain hoppings
will in general lead to instabilities towards some possible ordered
phases. The phase diagram of a system of coupled chains, including
ordered phases, was studied in Refs. \cite{Balentsa,Lin,Wu,Nickel_2005,Nickel_2006},
e.g. The FL behavior appears for energy scales higher then the critical
temperatures of such ordered phases. 

Moreover, for energies higher than some characteristic energy of the
order of the inter-chain hopping amplitude (possibly renormalized
by the interactions), one expects to recover LL features. Thus only
for intermediate energies is the FL picture expected to hold. Indeed
in Ref. \cite{Clarke} it was argued that even though the transverse
hopping is a relevant perturbation, in the RG sense, incoherent single
particle hopping between chains can lead to a LL-like behavior. This
was confirmed in \cite{Poilblanc,Capponi} using exact diagonalizations
(ED) and quantum Monte Carlo (QMC) since the incoherent part of the
spectral function (SF) is less affected by inter-chain hopping, and
the Drude weight is small compared to the incoherent weight, even
for small $\theta$. For larger $\theta$ the hopping between chains
becomes fully incoherent. Furthermore, considering a higher dimensional
mesh of coupled LL it was shown that there are mixed characteristics
of LL and FL \cite{Guinea}. 

A rather unifying picture was obtained using chain dynamical mean-field
theory (CDMFT)\cite{Biermann_2001a}. These studies observe a crossover
from a LL at high temperatures to a FL at low T with the coexistence
of a Drude feature with small spectral weight and a large incoherent
weight.

Several studies also treated the case of coupled Mott insulators.
The RPA approximation was used in \cite{Essler} and it was found
that for high enough hopping and small enough Coulomb coupling the
Mott-Hubbard gap closes and small Fermi pockets appear in the Fermi
surface with a finite $Z$. However, it was shown using CDMFT that
when the gap closes there is a continuous FS and no pockets \cite{Biermann_2001a}.
These results were also confirmed in \cite{Berthod} but it was found
that between the Mott phase and the FS phase there is an intermediate
phase where there are pockets (arcs because of spectral weight inhomogeneities).
Defining the FS both by the poles and zeros of the $\text{Re}G(\omega,\mathbf{k})$
it was shown that the Luttinger Theorem \cite{Dzyaloshinskii_2003}
is satisfied.

\vspace*{0.6cm}

The paper is structured as follows: In section \ref{sec:model} we
discuss the model and method used, briefly reviewing the RPA approach.
In section \ref{sec:Spectal} we present results for the spectral
function at low energies where a Luttinger-liquid-like universal description
holds and compare the results with other methods previously obtained.
In section \ref{sec:Spectral-function-at} we consider the regimes
of finite energies and consider finite but large $U$ values, the
infinite $U$ limit where the spins are dispersionless and the half-filing
Mott-insulator case. In section \ref{sec:frustration} we study the
role of frustration comparing a square, a triangular and a fully frustrated
lattices. We present some conclusions in section \ref{sec:Discussion}.
Also, in Appendix \ref{sec:Ap_restricted} we review the method of
Ref. \cite{Kohno_2007} developed for the spin structure factor of
the Heisenberg antiferromagnet in a triangular lattice and present
its generalization to the electron spectral function. In Appendix
\ref{sec:Ap_Factorized} we briefly review the method used for the
calculation of the spectral function for the Hubbard chain. In Appendix
\ref{sec:Ap_RPA} we review the derivation of the RPA formulation
and derive the expansion for its leading correction. This involves
the knowledge of higher correlation functions for the Hubbard chain,
which are not available at this time.

\section{Model and Method\label{sec:model}}

\begin{figure}[b]

\begin{centering}
\includegraphics[width=0.6\textwidth]{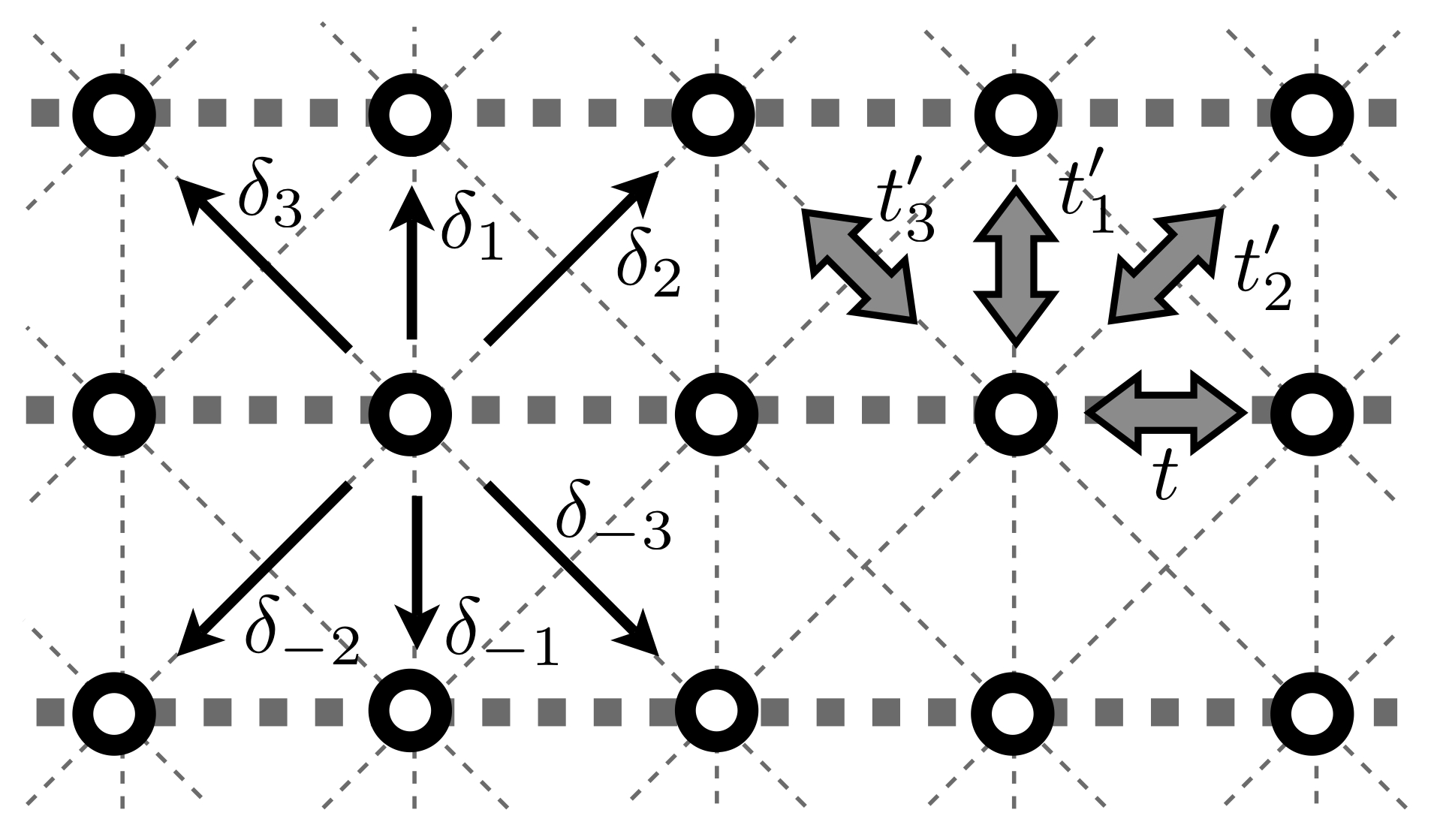} 
\par\end{centering}

\caption{\label{fig:t_tp} Direction of the hopping terms in the 2D Hubbard
model. }

\end{figure}

This section presents the method used to obtain the spectral function
of the weakly coupled Hubbard chains in terms of the one dimensional
spectral function. In order to set the notation we write the Hamiltonian
for the 2D Hubbard model as sum of an intra and an inter-chain term,
\begin{eqnarray*}
H & = & \sum_{y}H_{U,y}+H_{\perp},\end{eqnarray*}
 here \begin{eqnarray*}
H_{U,y} & = & \sum_{x,\sigma}-t\ \left[c_{x,y,\sigma}^{\dagger}c_{x+1,y,\sigma}+c_{x+1,y,\sigma}^{\dagger}c_{x,y,\sigma}\right]\\
 &  & +U\sum_{x,y}n_{x,y,\uparrow}n_{x,y,\downarrow}\end{eqnarray*}
 is the intra-chain contribution to the Hamiltonian of a chain parallel
to the $x$ direction and the subscript $y$ labels the direction
perpendicular to the chains. The hopping amplitude between the sites
in the chain is denoted by $t$, while $U$ is the usual on-site repulsion
that penalizes doubly occupancy of a given site. The transverse term
is given by \begin{align*}
H_{\perp} & =-\sum_{i}\sum_{\mathbf{r},\sigma}t'_{i}\ \left(c_{\mathbf{r},\sigma}^{\dagger}c_{\mathbf{r}+\boldsymbol{\delta}_{i},\sigma}+c_{\mathbf{r}+\boldsymbol{\delta}_{i},\sigma}^{\dagger}c_{\mathbf{r},\sigma}\right)\end{align*}
 where $t'_{i}$ labels the different inter-chain hoppings along the
directions $\boldsymbol{\delta}_{i}$ displayed in Fig.\ref{fig:t_tp}.
As shown, setting $t'_{2}=t'_{3}=0$ corresponds to an anisotropic
square lattice, $t'_{2}=t'_{1}$ and $t'_{3}=0$ to an anisotropic
triangular lattice, and $t'_{1}\ne0$ and $t'_{2}=t'_{3}$ to the
square lattice with diagonal hoppings.

\subsection{RPA and spectral function \label{sub:RPA-expression}}

As briefly reviewed in Appendix \ref{sec:Ap_RPA}, the single electron
Green's function in the so called random phase approximation (RPA)
is given by \begin{eqnarray}
G\left(\omega,\mathbf{k}\right) & = & \left[G_{1D}^{-1}\left(\omega,k_{x}\right)-t'\left(\mathbf{k}\right)\right]^{-1}\label{eq:RPA}\end{eqnarray}
 where \begin{eqnarray}
 &  & t'(\mathbf{k})=-2\sum_{i}t'_{i}\cos\left(\mathbf{k}.\boldsymbol{\delta}_{i}\right)\label{eq:t_k}\\
 &  & =-2\left[t'_{1}\cos\left(k_{y}\right)+t'_{2}\cos\left(k_{y}+k_{x}\right)+t'_{3}\cos\left(k_{y}-k_{x}\right)\right]\nonumber \end{eqnarray}
is the Fourier transform of the hopping matrix. Here $G_{1D}$ is
the Green's function of the one-dimensional system, assumed to be
known. In this work it is calculated exactly. The Fermi momentum $\mathbf{k}_{F}$
and the QP weight are obtain from Eq. (\ref{eq:RPA}) requiring\begin{eqnarray}
 &  & G^{-1}\left(\omega=0,\mathbf{k}_{F}\right)=0\label{eq:BoundStates}\\
 &  & Z^{-1}=\partial_{\omega}G^{-1}\left(\omega=0,\mathbf{k}_{F}\right).\label{eq:BoundStates-2}\end{eqnarray}
 In several works pioneered by Wen \cite{Wen_1990} this expression
has been used to study weakly coupled Luttinger Liquids \cite{Boies_1995,Tsvelik_1996}.
Note that Eq. (\ref{eq:RPA}) is exact for non-interacting electrons
$\left(\theta=0\right)$. 

Using Eq. (\ref{eq:RPA}) and the asymptotic form of the retarded
Green's function, in the low energy limit given by bosonization and
parameterized by \begin{equation}
\theta=\frac{1}{4}\left(K_{c}+\frac{1}{K_{c}}-2\right)\label{eq:theta}\end{equation}
 where $K_{c}$ is the Luttinger parameter, it was shown \cite{Wen_1990,Boies_1995,Tsvelik_1996,Gogolin_1998}
that for $\theta<1$ there is a nonvanishing QP weight \begin{equation}
Z\sim\left(\frac{v_{s}}{v_{c}}\right)^{\gamma}\left|\frac{t'\left(\mathbf{k}\right)}{\Lambda}\right|^{\frac{\theta}{1-\theta}}\label{eq:Z_CFT}\end{equation}
 for an arbitrary $t'\neq0$, where $\Lambda$ is an energy cutoff
and $\gamma$ is a exponent that can be explicitly computed (see chap.
19 of \cite{Gogolin_1998}). Note that for the non-interacting case
$\theta=0$, the low energy regime of the infinite $U$ limit of the
Hubbard model is recovered setting $\theta=1/8$ and higher values
of $\theta$ correspond to models with long range interaction. Besides
the region $\theta>1$, where no coherent mode is found at the RPA
level in \cite{Tsvelik_1996,Gogolin_1998}, the authors considered
the regimes $\theta<1/2$ and $\theta>1/2$ for which the exponent
$\gamma$ in (\ref{eq:Z_CFT}) changes from positive to negative.
They concluded that the value of the QP residue will be larger in
the second region. We will see further that there is a clear physical
signature separating these two regimes.

As stated in the introduction $t'$ is a relevant perturbation in
the RG sense, and thus the above treatment is valid only for energies
$T,\omega>T_{c}$ , where $T_{c}$ is the highest critical temperature
of all the possible order phases towards which the system is unstable
at low energy. Another energy scale is defined by $T_{1x}>T_{c}$
which separates a low energy regime where the pole of the Green's
functions is physically perceptible \cite{Boies_1995} from an higher
energy regime for which fully coherent 2D hopping is suppressed. For
the non-interacting case $\left(\theta=0\right)$ $T_{1x}$ is of
the order of the interchain coupling $t'$. It has been shown \cite{Boies_1995}
that for the interacting case this energy scale is reduced yielding
$T_{1x}\sim\Lambda\left(\frac{t'}{\Lambda}\right)^{1/\left(1-\theta\right)}$
for $\theta<1$. For $\theta>1$ this treatment leads to a vanishing
$T_{1x}$; however, as also noticed in \cite{Boies_1995}, higher
order terms that consider two-particle processes define another energy
scale $T_{2x}$ that will overtake $T_{1x}$ and further extend beyond
$\theta=1$ the region where $Z\neq0$. Note that these works are
only valid for arbitrarily small energies since the one dimensional
quantities are given by bosonization and thus no predictions can be
obtained for the moderate and high energy regimes. One of the aspects
of the present work is precisely to be able to access these regions.

Another feature of the RPA expression is that it leads to an anisotropic
QP weight along the FS which vanishes for $t'(\mathbf{k})=0$. This
could suggest the existence of hot-spots in the FS where the 1D character
would be strongly manifested. However, subsequent works, using exact
resummation of some infinite class of diagrams \cite{Arrigoni_1998,Arrigoni_1999,Arrigoni_2000}
and higher dimensional bosonization \cite{Kopietz_1997} pointed out
that the vanishing $Z$ was an artifact of the RPA and that the inclusion
of higher order terms leads to a smoothly varying QP along the FS;
this fact was also verified by DMFT calculations \cite{Georges_2000,Biermann_2001,Biermann_2001-b,Giamarchi_2004}.
All these works predict a finite QP pole leading to FL like behavior
for non-zero $t'$ for the Hubbard model.

However, the RPA expression gives a qualitative description of the
crossover from 1 to 2D. In practice the use of RPA-like expressions
has gathered a great success describing antiferromagnetic spin chains
\cite{Kohno_2007,Kohno_2009} whith a good quantitative agreement
with experiments. In electronic systems the DMFT approach, based in
a large $D_{\perp}$ (dimensionality of the transverse dimension)
expansion, obtained a good agreement for the frequency dependent interchain
conductivity \cite{Georges_2000}.

\begin{figure*}
\begin{centering}
\begin{tabular}{ccc}
\includegraphics[width=0.3\textwidth]{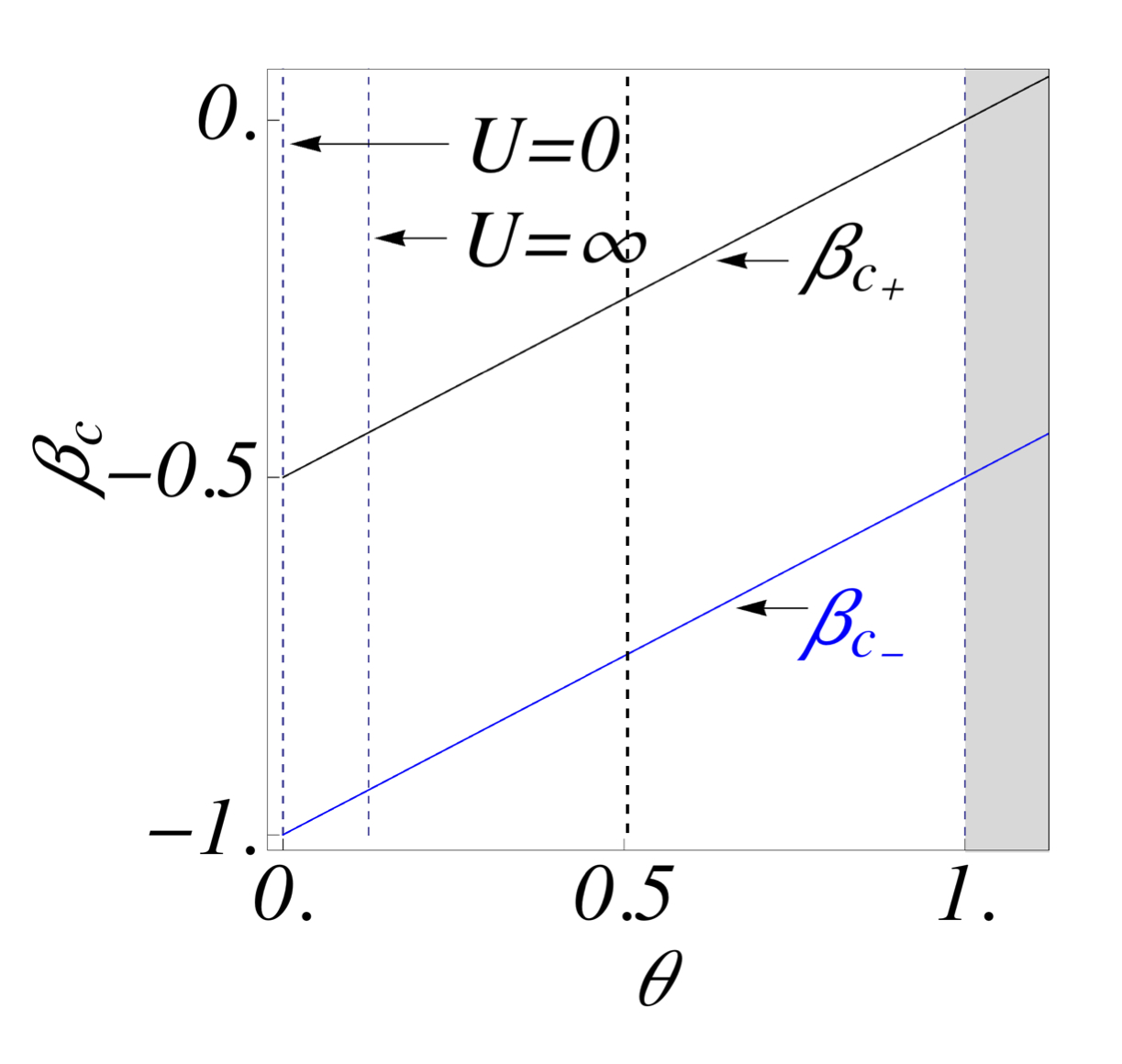}  & \includegraphics[width=0.3\textwidth]{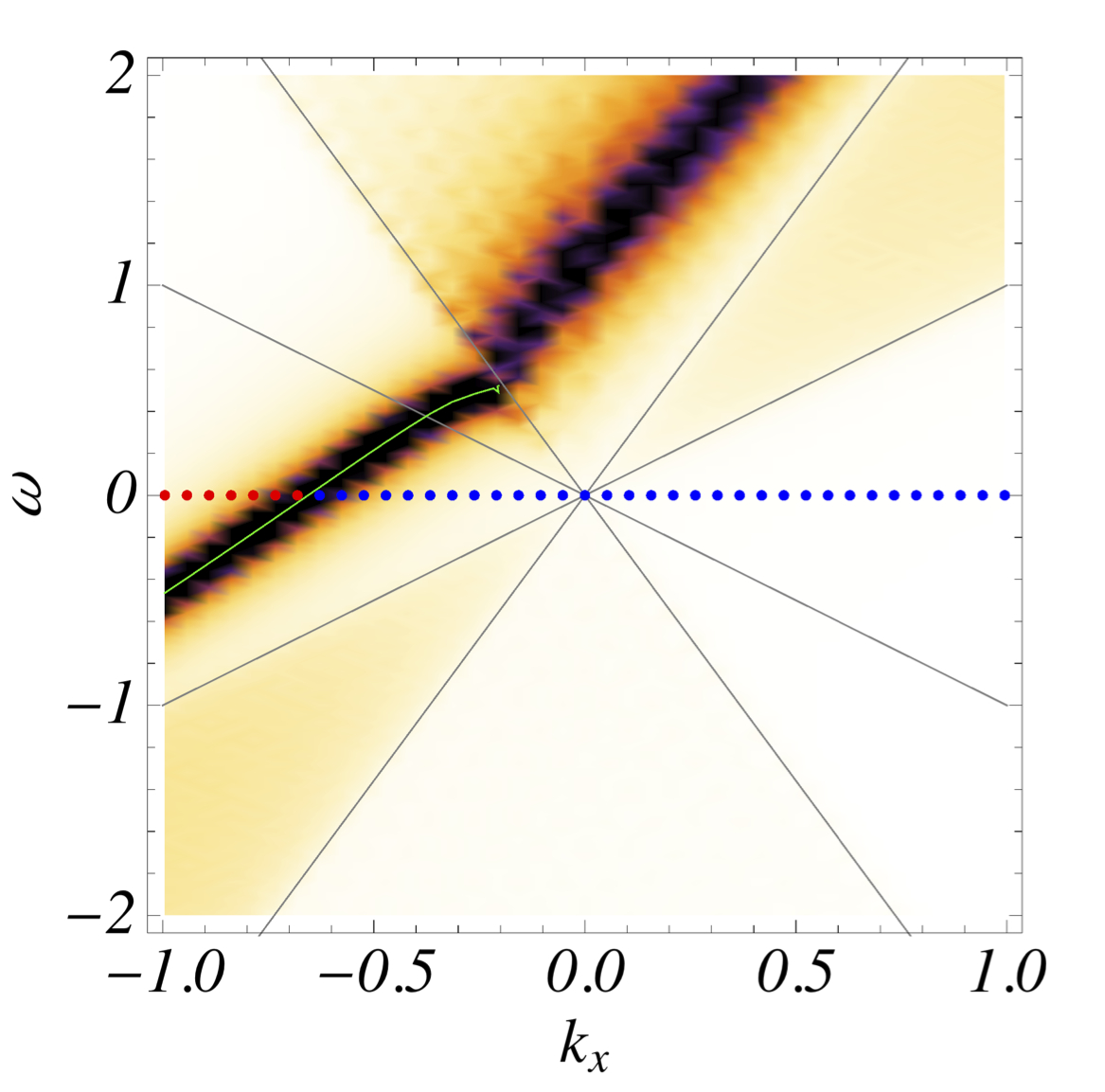}  & \includegraphics[width=0.3\textwidth]{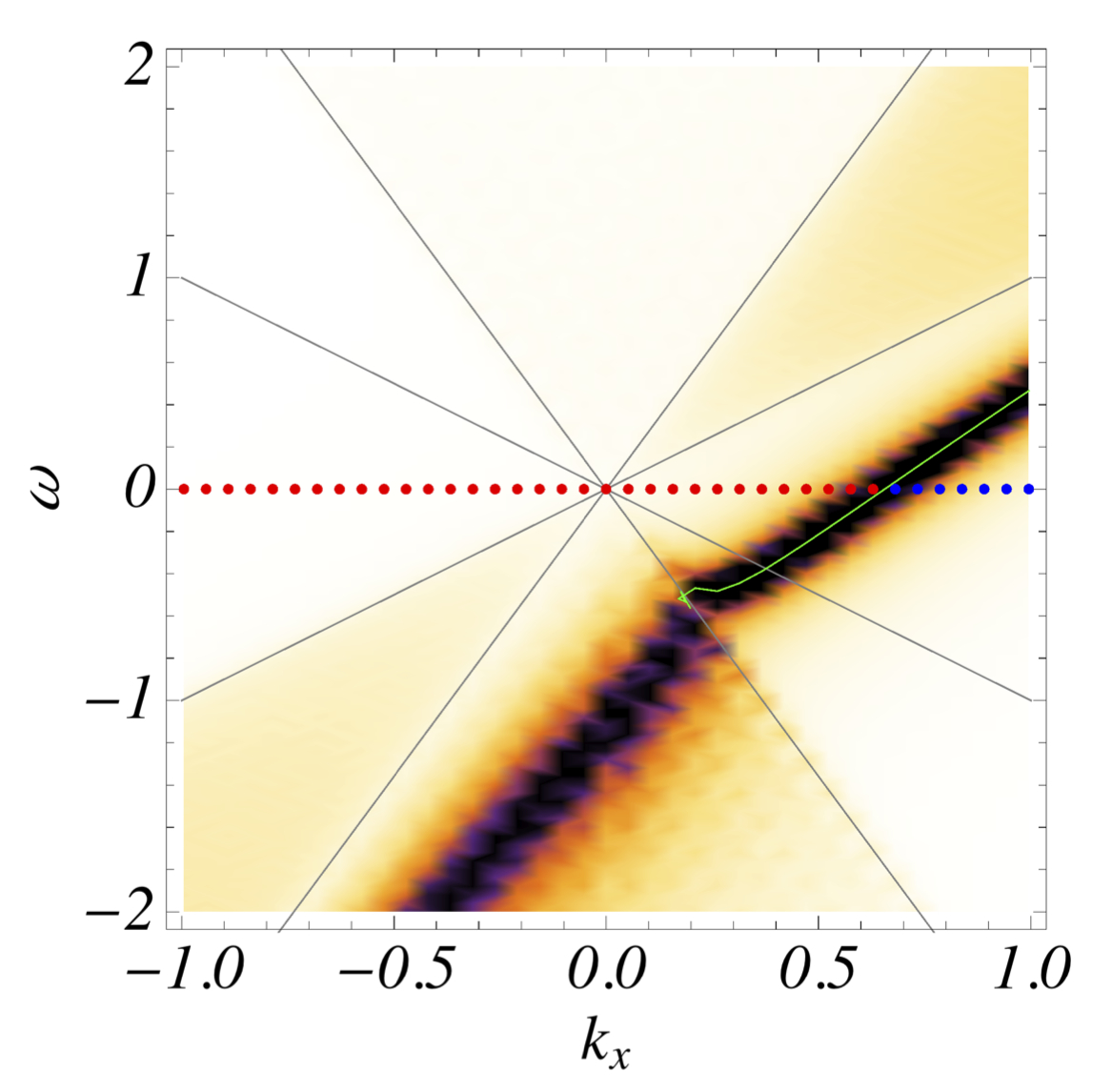}\tabularnewline
\end{tabular}
\par\end{centering}

\caption{\label{fig:Spectral_theta-0}\textbf{Left Panel: }Values of the charge
exponents $\beta_{c}^{+}$and $\beta_{c}^{-}$ as a function of the
LL parameter $\theta$. \textbf{Central and Right Panels:} Spectral
function obtained using the RPA expression (\ref{eq:RPA}) for $t'=0.2$
and $t'=-0.2$ respectively and for $\theta=1/8$, $v_{s}=1$ and
$v_{c}=2.718$. The Gray lines signal the boundaries of spin and charge
continuum of the 1D spectral function: $\omega=\pm u_{s}k$ and $\omega=\pm u_{c}k$.
The bound state (Green line) is obtained solving Eq. (\ref{eq:BoundStates})
. Red and Blue dots displayed at $\omega=0$ correspond to values
of $k_{x}$ for which $\sign\re\left[G\left(\omega=0,k_{x}\right)\right]$
is respectively positive or negative.}

\end{figure*}

In this work we use two approaches to compute the one dimensional
spectral function. The first is only valid for low energies and is
equivalent to the use of the asymptotic Green's function given by
bosonization. It was used to verify the predictions referred in the
last section and to understand the low energy limit of the second
approach valid for arbitrary energies. Due to its generality it permits
to vary independently the interaction strength (changing $\theta$)
as well as the spin and charge velocities. The second approach relies
on the exact solution of the large $U$ limit of the Hubbard model.
This limit permits considerable simplifications and in particular
a closed form for the 1D spectral function. A detailed description
of both methods in given in the following sections.

The lowest lying excitations contributing to the spectral function
of the 1D Hubbard model correspond to the creation of a holon and
a spinon (charge and spin excitations). These two quasiparticles propagate
with different velocities and in terms of the original electrons are
very complex. Even though they have a fractionalized existence inside
the 1D many-body system, when an electron is, for instance, removed
from the chain (photoemission) they recombine. If the chains are weakly
coupled one expects that the excitations travel along the transverse
direction as \textquotedbl{}electrons\textquotedbl{}. The holon and
the spinon are expected to propagate coherently from one chain to
the next. This idea was proposed in Ref. \cite{Kohno_2007} in the
context of an antiferromagnet in a triangular lattice. In the 1D Heisenberg
antiferromagnet the low lying excitations are two spinons. In the
weak coupling regime they are assumed to propagate coherently from
chain to chain (like a $\Delta S=1$ excitation -- a magnon). It is
therefore interesting to generalize the procedure developed in Ref.
\cite{Kohno_2007}, for the spin structure factor of the antiferromagnetic
Heisenberg model, to the present case of the spectral function of
the Hubbard model. This is carried out in Appendix \ref{sec:Ap_restricted}.
There are however difficulties associated with instabilities of the
system resulting from the approximation used. The expression obtained
for the spectral function is formally very similar to the one obtained
within RPA (Appendix \ref{sec:Ap_RPA}) as noted in ref. \cite{Kohno_2007}
for the antiferromagnet. A basic difference is that in the RPA the
spectral function is defined as a complete function (for positive
and negative energies) while in the restricted Hilbert space considered
in Appendix \ref{sec:Ap_restricted}, the positive and negative energies
are associated with two functions defined separately. Due to the appearance
of bound states one is led to a situation where the excited states
have negative energies, which implies an instability of the groundstate.
Therefore we will use in the following the RPA expression (\ref{eq:RPA})
to obtain the 2D Green's function. In this context the bound states
are interpreted as coherent modes resulting from spectral weight transfer
among different energies, as discussed next.

To fix the notation we define the spectral function as \begin{equation}
Sp\left(\omega,\mathbf{k}\right)=-\frac{1}{\pi}\im G\left(\omega,\mathbf{k}\right).\end{equation}
 In the literature it is usual to write this quantity as a sum $Sp\left(\omega,\mathbf{k}\right)=A(\omega,\mathbf{k})+B(\omega,\mathbf{k})$
where \begin{equation}
A(\omega,\mathbf{k})=\sum_{f,\sigma}\left|\bra{f,N+1}c_{k,\sigma}^{\dagger}\ket{0,N}\right|^{2}\delta(\omega-E_{f}^{N+1}+E_{0}^{N})\end{equation}
 is the measured amplitude of angular resolved inverse photoemission
experiments, here given in the Lehmann representation, and \begin{equation}
B(\omega,\mathbf{k})=\sum_{f,\sigma}\left|\bra{f,N-1}c_{k,\sigma}\ket{0,N}\right|^{2}\delta(\omega-E_{0}^{N}-E_{f}^{N-1})\end{equation}
 the measured angular resolved photoemission amplitude. $N$ is the
number of electrons, $0$ and$f$ denote the ground and final states
respectively, the chemical potential is taken such that the ground
state corresponds to zero energy so $A(\omega<0,\mathbf{k})=0$ and
$B(\omega>0,\mathbf{k})=0$. \begin{align*}
\\\end{align*}

\section{Spectral function at low energies: Luttinger-Liquid-like regime \label{sec:Spectal}}

In this section we concentrate on the low energy region that is characterized
by linearized dispersions and power-law behavior, and study how the
2D spectral properties for low energies emerge as a function of $t'$
and $\theta$ within the RPA (\ref{eq:RPA}). We recover some results
by other authors, reviewed in the last section, and find some new
features characterizing the different regimes.

\begin{figure*}[!t]

\begin{centering}
\includegraphics[width=0.9\textwidth]{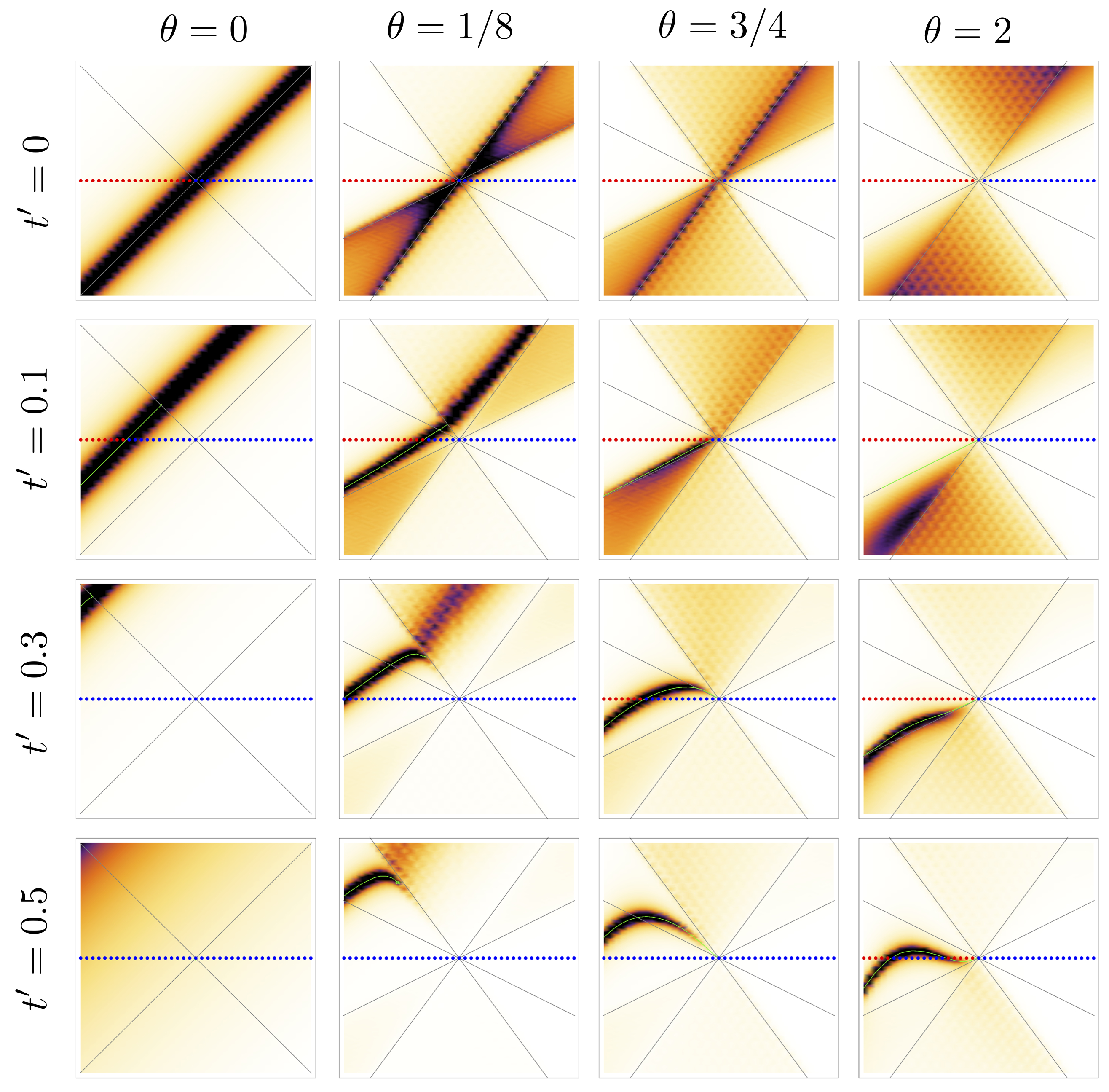} 
\par\end{centering}

\caption{\label{fig:Spectral_theta-1}Spectral function obtained by RPA formula
(\ref{eq:RPA}) for different values of the LL parameter $\theta$
and inter-chain coupling $t'$. Axes and labels are the same as for
Figs. \ref{fig:Spectral_theta-0} - Central and Right Panels. For
$\theta=0$ we set $v_{c}=v_{s}$ in order to obtain the exact free
particle result; for all other values of $\theta$ $v_{s}=1$ and
$v_{c}=2.718$. The Gray lines signal the boundaries of spin and charge
continuum. Red and Blue dots displayed at $\omega=0$ correspond to
values of $k_{x}$ for which $\sign\re\left[G\left(\omega=0,k_{x}\right)\right]$
is respectively positive or negative. For the cases where $Z\neq0$
this criterion corresponds to $k_{x}$ being inside or outside the
Fermi surface. The Green line corresponds to the bound states obtained
solving Eq. (\ref{eq:BoundStates}). }

\end{figure*}

For low energies, and near the Fermi momentum, the spectral function
of one dimensional gapless electronic systems can be written as a
convolution of the spin and charge parts \begin{eqnarray}
Sp\left(\omega,k\right) & \propto & \sum_{i,j;i',j'\in\mathbb{N}}a_{i,j}^{c}a_{i',j'}^{s}\left[\delta\left(\omega-\Omega_{i,j}^{c}-\Omega_{i',j'}^{s}\right)\delta_{k,K_{i,j}+K_{i',j'}}+\delta\left(\omega+\Omega_{i,j}^{c}+\Omega_{i',j'}^{s}\right)\delta_{k,-K_{i,j}-K_{i',j'}}\right]\label{eq:SF_CFT}\end{eqnarray}
 where $K_{i,j}=2\pi(i-j)/L$ are the momenta of the excitations,
$\Omega_{i,j}^{\alpha}=2\pi v_{\alpha}(i+j)/L$ are the corresponding
energies (with $\alpha=c,s$ and $v_{s}$ and $v_{c}$ are the spin
and charge velocities) and their weights are explicitly given by \begin{eqnarray}
a_{i,j}^{\alpha} & = & \frac{\Gamma\left(i+\beta_{\alpha}^{+}+1\right)}{i!\,\Gamma\left(\beta_{\alpha}^{+}+1\right)}\frac{\Gamma\left(j+\beta_{\alpha}^{-}+1\right)}{j!\,\Gamma\left(\beta_{\alpha}^{-}+1\right)}\;.\end{eqnarray}
 The exponents $\beta_{c}^{+},\beta_{c}^{-}$ and $\beta_{s}^{+},\beta_{s}^{-}$
characterize the divergence of the spectral function at the edges
of the (right, $+$ and left, $-$) charge and spin continua at either
the right or left Fermi points. For a Luttinger liquid with SU(2)
spin rotation symmetry both $\left\{ \beta_{s}^{+},\beta_{s}^{-}\right\} =\left\{ -\frac{1}{2},-1\right\} $
are fixed. The charge exponents are given by \begin{eqnarray}
\left\{ \beta_{c}^{+},\beta_{c}^{-}\right\}  & = & \left\{ \frac{\theta}{2}-\frac{1}{2},\frac{\theta}{2}-1\right\} ,\end{eqnarray}
 where $\theta$ is related with the Luttinger parameter $K_{c}$
by Eq. (\ref{eq:theta}) (see also Fig.\ref{fig:Spectral_theta-0}).
As we already mentioned, $\theta=0$ for the noninteracting fermions,
and $\theta\rightarrow1/8$ for $U\rightarrow+\infty$ Hubbard model.

The particular form of the spectral function given by Eq. (\ref{eq:SF_CFT})
was obtained in Ref. \cite{Penc_1995} in the context of the large
$U$ approximation of the Hubbard model. However, the described low
energy structure is much more general and can be traced back to the
conformal invariance of the (1+1)D model \cite{Cardy_1986}. In the
thermodynamic limit one obtains the well known asymptotic form, say
for the right moving electrons, of the real time Green's function
\begin{eqnarray*}
G_{r}\left(x,t>0\right) & \simeq & \frac{e^{ik_{F}x}}{(x-v_{c}t)^{1+\beta_{c}^{+}}(x+v_{c}t)^{1+\beta_{c}^{-}}}\times\\
 &  & \frac{1}{(x-v_{s}t)^{1+\beta_{s}^{+}}(x+v_{s}t)^{1+\beta_{s}^{-}}}\end{eqnarray*}
 that can be directly obtained by bosonization techniques.

\begin{figure}
\begin{centering}
\includegraphics[width=0.5\textwidth]{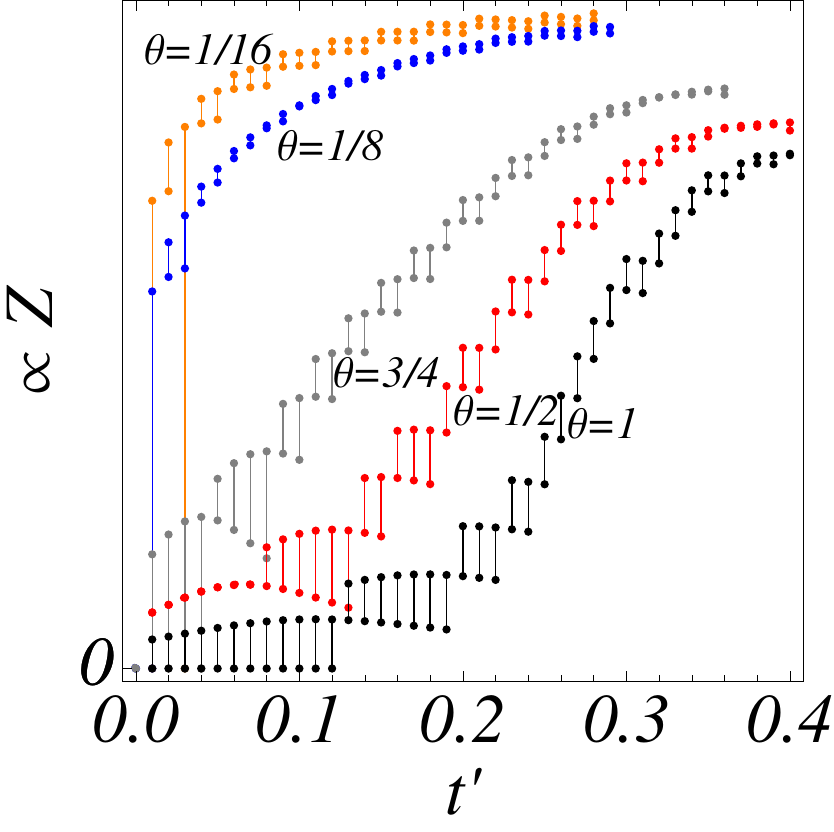} 
\par\end{centering}

\caption{\label{fig:Spectral_theta-2}Quasiparticle residue as a function of
$t'$ for different values of $\theta$: $\theta=1/16$ (Orange),
$\theta=1/8$ (Blue), $\theta=1/2$ (Gray) and $\theta=1$ (Black).
For each value of $t'$, $k_{x}^{+}$ and $k_{x}^{-}$ were determined
on each side of the FS. For these values the bound state equation
was solved in order to find $\omega_{+/-}\simeq0$, $Z_{+/-}$ was
computed using Eq. (\ref{eq:BoundStates-2}). }

\end{figure}

With the 1D Green's function computed with (\ref{eq:SF_CFT}) we used
Eq. (\ref{eq:RPA}) to obtain the 2D spectral function $Sp\left(\omega,\mathbf{k}\right)=-\frac{1}{\pi}\im G\left(\omega,\mathbf{k}\right)$
for a fixed value of $t'(\mathbf{k})=t'$. In Figure \ref{fig:Spectral_theta-0}
- (Central and Right panels) we show the typical results obtained
here. The bound states were found solving Eq. (\ref{eq:BoundStates})
outside the spin and charge continua. The FS was determined for the
values of $k$ for which $\re G\left(\omega,\mathbf{k}\right)$ changes
sign. Figure \ref{fig:Spectral_theta-1} displays the main results
of this section. The spectral function is shown for different values
of the LL parameter $\theta$ and inter-chain coupling $t'$. For
$\theta=0$ we set $v_{c}=v_{s}$ in order to obtain the exact free
particle result; for all other values of $\theta$, fixed values of
the spin and charge velocities were used for the physical case $v_{s}<v_{c}$.
Fig. \ref{fig:Spectral_theta-2} shows the values of the QP residue
as a function of $t'$ for different values of $\theta$. The error
bars are due to the discreteness of the $k$ values: for each value
of $t'$, $k_{x}^{+}$ and $k_{x}^{-}$ were determined on each side
of the FS. For these values the bound state equation was solved in
order to find $\omega_{+/-}\simeq0$; $Z_{+/-}$ was then computed
using Eq. (\ref{eq:BoundStates-2}).

As a general feature, we note the change from incoherent regions,
arising from the spin-charge separation in the 1D case ($t'=0$),
to the sharply defined coherent excitations as $t'$ increases. For
the 1D case the spectral function is strictly zero outside the 1D
continuum, delimited by the spin and charge velocities (see \ref{fig:Spectral_theta-1}
upper-left panel). The interchain coupling $t'$ favors the appearance
of sharp coherent features not only outside the 1D continuum, where
they correspond to poles of the 2D Green's function, but also within
the 1D continuum where the spectral weight also tends to concentrate.
For $\theta<1$ and small positive $t'$($\simeq0.1$) there is a
considerable transfer of spectral weight to the spin (charge) branches
for $\omega<0$ ($\omega>0$) . For negative $t'$ the spin and charge
roles are interchanged (see Fig. \ref{fig:Spectral_theta-0} - Central
and Right Panels). The critical value of $\theta=1$, predicted by
several authors \cite{Wen_1990,Boies_1995,Tsvelik_1996,Gogolin_1998},
is found such that for $\theta<1$ a bound state appears for $t'\neq0$
crossing $\omega=0$ at $k_{F}(t')\neq k_{F}(t'=0)$ changing the
position of the Fermi surface and resulting in a non-vanishing QP
weight $Z$. For $\theta>1$ Fig. \ref{fig:Spectral_theta-1} shows
that for small values of $t'$ the bound state still appears. However,
since it does not cross $\omega=0$, it is unable to drive the system
to a FL like behavior. After some critical $t'$ is reached the bound
state crosses twice the $\omega=0$ line creating a hole pocket. Note,
however, that in this regime the RPA is probably out of its domain
of validity and this last feature is probably an artifact. In figure
Fig. \ref{fig:Spectral_theta-1} we show the evolution of the quasiparticle
residue as a function of theta. For the region $\theta<1/2$ a damped
mode is observed when the coherent mode enters the charge continuum.
For $\theta>1/2$ this feature disappears and the coherent mode is
deflected to $\omega=0$ and loses all its spectral weight before
entering in the continuum. This feature clearly differentiates both
regimes. The QP residue as a function of $t'$ is shown in Fig. \ref{fig:Spectral_theta-2}.
The large error bars obtained due to the discreteness of the values
of $k$ prevent a clear fit.

\section{Spectral function at finite energies\label{sec:Spectral-function-at}}

\begin{figure}
\begin{centering}
\includegraphics[width=0.5\textwidth]{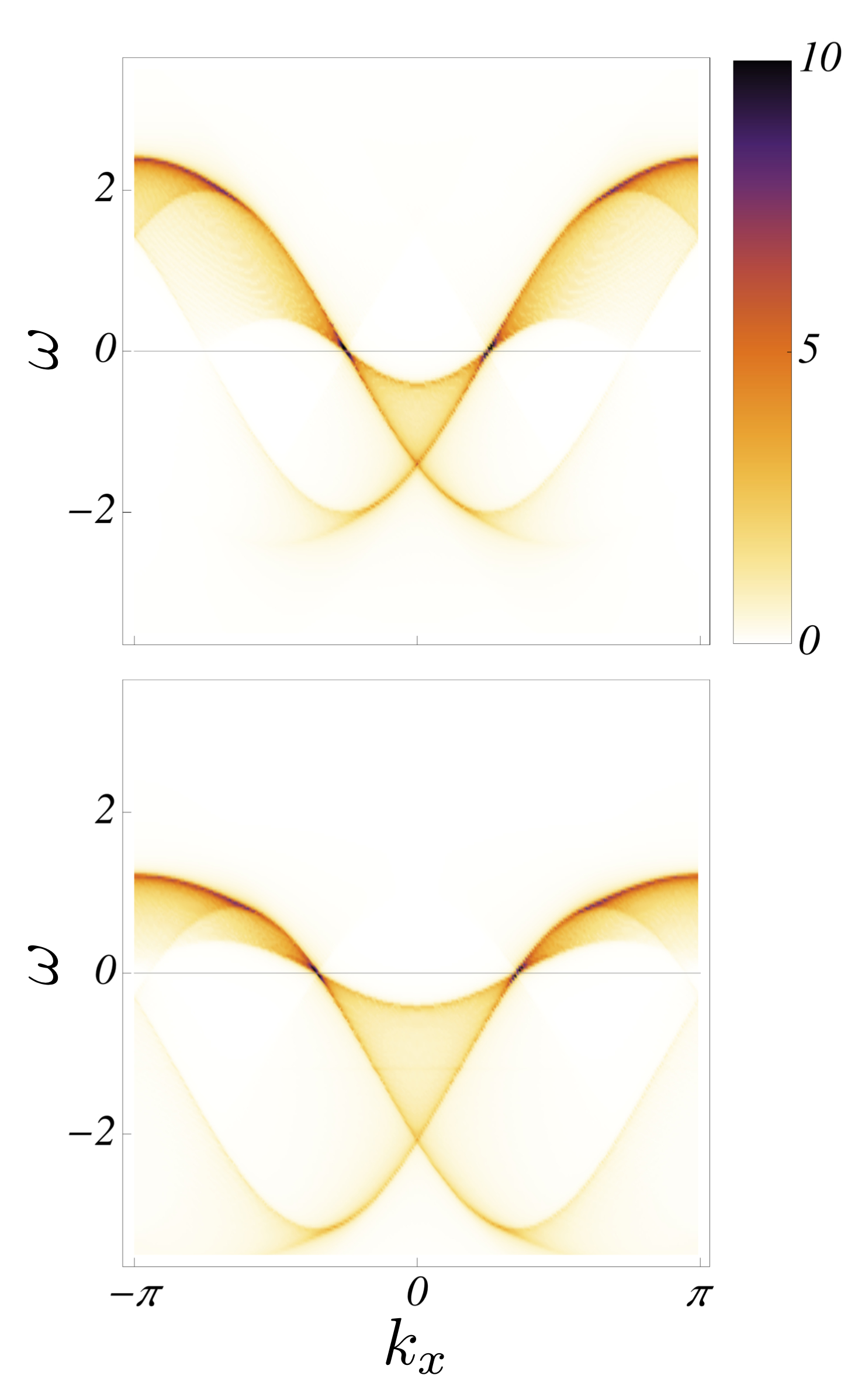} 
\par\end{centering}

\caption{\label{fig:Spectral_HC_np5_1}Spectral function for the one dimensional
Hubbard Model within the large $U$ approximation. Upper Panel: quarter-filling
$n=0.5$, computed for $U/t=7.5$. Lower Panel: $n=0.7$, computed
for $U/t=11.5$ }

\end{figure}

In this section we use the spectral function obtained for large $U$
\cite{Penc_1997}, together with the RPA expression (\ref{eq:RPA}),
to compute the finite energy spectral function, for weakly coupled
Hubbard chains. The results presented here generalize to finite energies
the ones obtained in the previous section for systems that can be
well described by an Hubbard like Hamiltonian, with relatively large
onsite repulsion ($U/t\agt6$).

It has been shown that in the $U\to\infty$ limit the eigenstates
of the Hubbard chain can be written as a product of a spinless free
fermion and a squeezed spin wave functions \cite{Woynarovich82,Ogata_1990}.
In subsequent works \cite{Penc_1995,Penc_1997} this factorized form
was used to write the spectral function as a convolution over the
spin and the fermionic parts (see Appendix \ref{Sec:Appendix:factorizedWF}).
The nontrivial fermionic matrix elements are computed between wave
functions of free fermionic states on a ring, with different twisted
boundary conditions imposed by the spin configurations. This simplification
permitted to obtain the spectral function in the infinite $U$ limit.
Note however that if $U\to\infty$ the spin spectrum collapses and
the spin sector is completely degenerate.

Once the $t/U$ is finite, the problem can be treated perturbatively,
and to get the first order corrections of the energy it is sufficient
to look at the expectation value of the perturbing Hamiltonian ($\propto t/U$)
with the unperturbed, spin-charge factorized wave functions. When
calculating the spectral functions, additional corrections appear
in the matrix elements that come from applying the unitary transformation
to the electron creation and anihilation operators \cite{harris,oles}.
For our purposes the most important effect of the finite $t/U$ is
to introduce a finite spinon velocity, and that is already captured
by the first order corrections to the energy. The spinon velocity
at the Fermi momenta is given by \begin{eqnarray}
v_{s} & = & \frac{2\pi t^{2}}{U}\left(1-\frac{\sin2\pi n}{2\pi n}\right)+O\left(\frac{1}{U^{2}}\right),\label{eq:u_s_of_U}\end{eqnarray}
where $n$ is the band-filling, and the exponents are calculated at
the Fermi level.

The results of \cite{Penc_1997} and its extension to finite $U$
were proven to be quite accurate for $U/t\agt6$ (see \cite{Carmelo_2006-1}).
Using this method the 1D spectral function was obtained considering
systems with size $L$, ranging typically from 120 to 300 sites; quantitative
differences as a function of $L$ were observed to be small within
this range. Moreover, in order to reduce the computational time, the
results presented here used only contributions from one and two particle-hole
excitations that were shown to carry the vast majority of the spectral
weight ($>99\%$) \cite{Penc_1997}; the inclusion of higher order
processes was observed to give neglectable contributions. The values
of $U$ were obtained fitting the spin velocity $v_{s}$ with the
expression (\ref{eq:u_s_of_U}), after having computed the 1D spectral
function with an effective exchange constant $J_{\text{eff}}$ of
the order of $\simeq0.2$. Using the RPA expression (\ref{eq:RPA}),
the 2D spectral function was computed for different values of the
band filling and transverse momentum. The exact position of the bound
state dispersion was obtained as well as the new FS and the dependence
of the QP weight. The results are presented in the next sections.

\subsection{Finite large U}

\begin{figure*}
\begin{centering}
\includegraphics[width=0.9\textwidth]{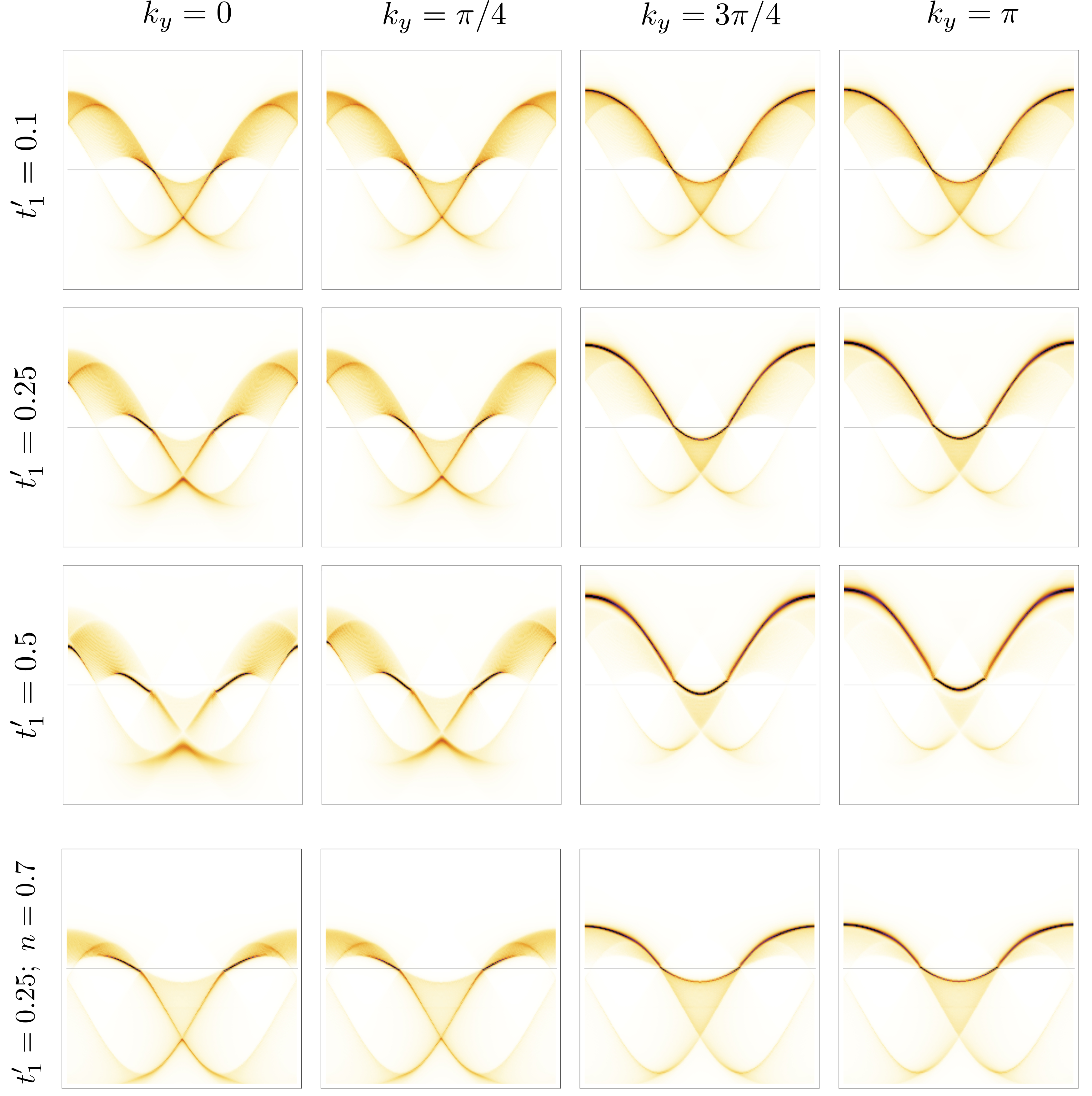} 
\par\end{centering}

\caption{\label{fig:Spectral_HC_np5_2} Top 3 rows: Spectral function of the
Hubbard model at quarter-filling ($n=0.5$) and $U=7.5$ in an anisotropic
square-lattice obtained by weakly coupling Hubbard chains within the
RPA (\ref{eq:RPA}) for different values of the inter-chain hopping
$t'_{1}$ and transverse momentum $k_{y}$. The axes labels and the
scale are the same as for Figs. \ref{fig:Spectral_HC_np5_1} . As
$t'_{1}$ increases the bound states, corresponding to a coherent
excitation, start on the boundaries of the continuous region changing
the shape of the FS. Lower row: Spectral function of the Hubbard model
for $n=0.7$ and $t'_{1}=0.25$ and $U/t=11.5$.}

\end{figure*}

\begin{figure}
\begin{centering}
\includegraphics[width=0.7\textwidth]{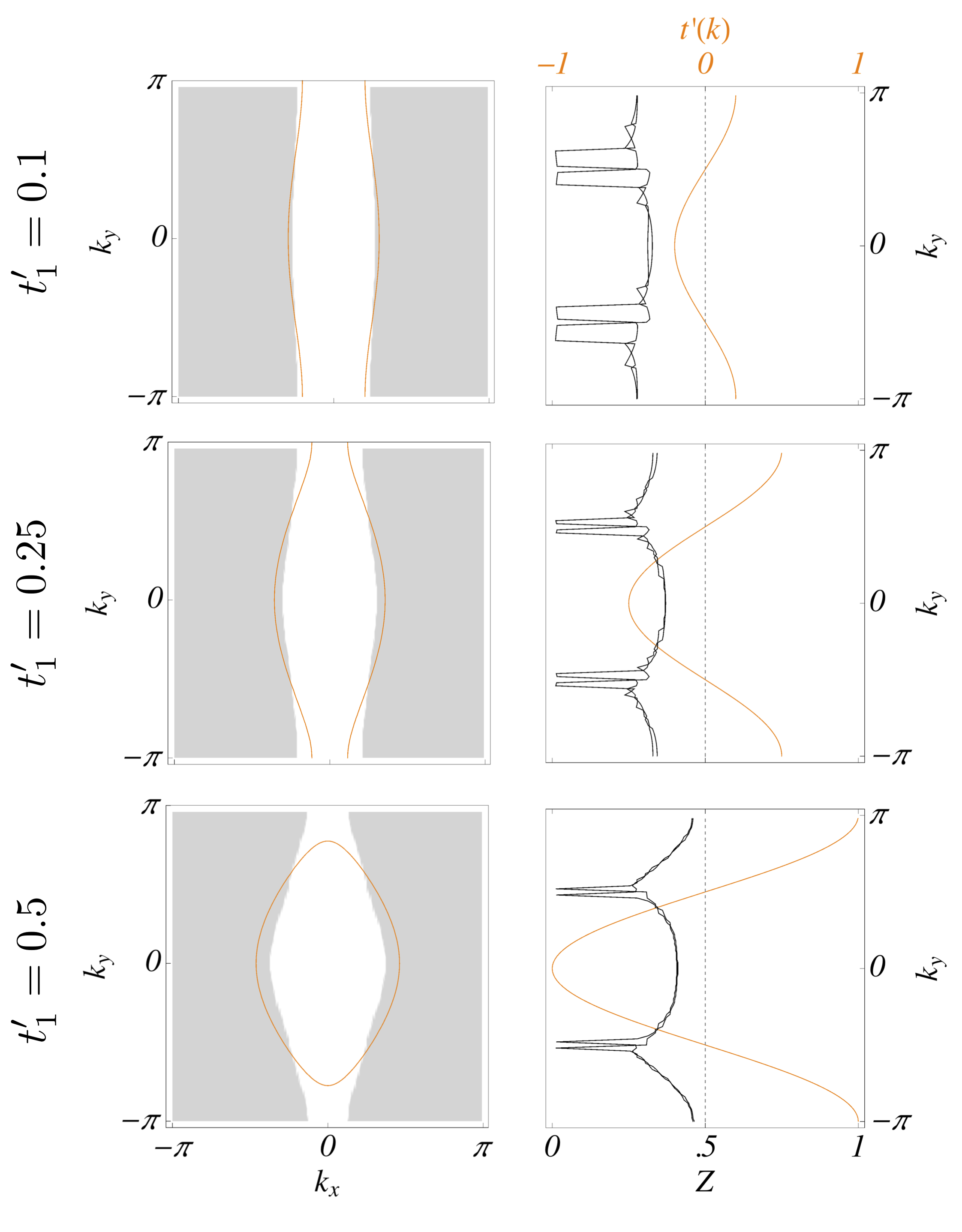} 
\par\end{centering}

\caption{\label{fig:Spectral_HC_np5_3}Left Panel: The Gray and White regions
correspond, respectively, to $\sign\re\left[G\left(\omega=0,k_{x}\right)\right]<0$
and $\sign\re\left[G\left(\omega=0,k_{x}\right)\right]>0$ , i.e.
to the exterior and interior of the FS obtained within the RPA. The
FS of non-interacting fermions is given by the Orange curves. Right
Panel: Quasiparticle residue along the FS as function of $k_{y}$.
For each value of $k_{y}$, $k_{x}^{+}$ and $k_{x}^{-}$ were determined
on each side of the FS. For these values, the bound-state equation
was solved in order to find $\omega_{+/-}\simeq0$; $Z_{+/-}$ (Black
lines) was computed using Eq. (\ref{eq:BoundStates-2}). The RPA $t'\left(\mathbf{k}\right)$
along the FS is plotted as a function of $k_{y}$ (Orange curve).
As implied by the RPA expression, when the self energy vanishes the
coherent excitations disappear ($Z=0$). Upper, Central and Lower
panels correspond respectively to $t'_{1}=0.1;\,0.2;\,0.5$ and to
$n=0.5$. }

\end{figure}

Fig. \ref{fig:Spectral_HC_np5_1} shows the Hubbard chain ($t'=0$)
spectral function for quarter-filling $(n=0.5)$ and a large value
of $U/t=7.5$ and for $n=0.7;$ $U/t=11.5$. Close to zero energy
(chemical potential) there is a large spectral weight along both the
spinon and holon branch lines. Note that the spectral weight along
the spinon branch dies out as we move away from the Fermi level towards
positive energies, while the spectral weight along the holon branch
line remains high. The branch lines for arbitrary values of the Hubbard
coupling, $U$, are obtained moving one excitation (spinon or holon)
along its band while keeping the other one fixed at the Fermi level.
In the vicinity of the branch line the spectral weight has a power
law behavior with exponents that may be negative (yielding a large
spectral weight) or positive (leading to an edge and small spectral
weight). As shown in Fig. 1 of Ref. \cite{Carmelo_2004_b} the exponent
along the spinon branch line for positive energies changes sign from
negative to positive and, therefore, there is a loss of spectral weight,
while the exponent along the holon line is always negative.

The 1D results of Fig. \ref{fig:Spectral_HC_np5_1} are to be compared
with those of Fig. \ref{fig:Spectral_HC_np5_2} where the spectral
function is computed for an anisotropic square lattice $(t'_{2}=t'_{3}=0)$,
for different values of the interchain coupling and transverse momentum
and for band-fillings $n=0.5,0.7$. Note that since, in this approximation,
$t'(\mathbf{k})=0$ for $k_{y}=\pi/2$, the SF for this value of the
transverse momentum is given by the 1D case of Fig. \ref{fig:Spectral_HC_np5_1}.
The low energy behavior near the 1D Fermi momentum agrees with the
results of the previous section. In the left panels the effective
hopping $t'(k)>0$ and in the right panels the effective hopping is
$t'(k)<0$. As shown in the previous section this implies that for
$k_{y}<\pi/2$ the FS increases in size and for $k_{y}>\pi/2$ the
FS shrinks. In the 1D case there is a high spectral weight along both
the spinon and holon branches at the Fermi surface. Introducing the
transverse hopping we find as for the coupled LL that there is an
increased spectral weight in one of the two branches depending on
the sign of $t'(\mathbf{k})$: for $t'(k)>0$ at positive energies
the weight is concentrated in the spinon branch and at negative energies
in the holon branch while the opposite occurs for $t'(k)<0$.

Bound states arise near the spinon branch and their weight increases
with $t'$. For the low energy region the spectral weight outside
the 1D continuum is strictly zero due to phase space constraints.
In this case the sharp coherent features are poles of the 2D Green's
function. Besides the bound states near $\omega=0$, anti-bound states
are formed at high energies. However for the high energy part of the
1D spectrum there is generically no region with strictly zero spectral
weight since small contributions will come from higher order particle-hole
processes not considered in our method. This means that in practice,
contrarily to the boundstates, anti-boundstates will have a small
width corresponding to a long, but finite, lifetime of this QP-like
excitations. As the transverse hopping increases, all the coherent
features become sharper inside and outside the 1D continuum. However,
there is still a significant distribution of spectral weight through
a continuum, a 1D characteristic. Note that a bound state emerges
from the edge of the Brillouin zone that extends to lower energies,
as the transverse hopping grows.

In Figure \ref{fig:Spectral_HC_np5_3} we show the 2D Fermi surface
and the quasiparticle residues for different transverse hoppings.
The left panels of Fig. \ref{fig:Spectral_HC_np5_3} show the evolution
of the FS as the interchain hopping is increased. Comparison with
the non-interacting case (Orange line) shows that interactions will
tend to prevent warping of the FS. The Right Panels display the value
of $Z$ (black lines) and $t'(\mathbf{k})$ (orange lines) along the
FS. The QP weight clearly increases with $t'_{1}$. Along the FS the
inhomogeneities of $Z$ are quite smooth except for the vicinity of
$k_{y}=\pi/2$ where it vanishes. As discussed in sec. \ref{sub:RPA-expression},
since $t'(\mathbf{k})=0$ at this point the RPA expression is known
to fail. Higher order corrections will give a finite $Z$ value leading
to a non-zero QP weight along the FS and thus to FL-like behavior.
Note also that even for $t'(\mathbf{k})\neq0$ the RPA underestimates
the value of $Z$, so higher order corrections will be expected to
slightly increase its value.

\subsection{Infinite U }

\begin{figure*}[!t]

\begin{centering}
\includegraphics[width=0.9\textwidth]{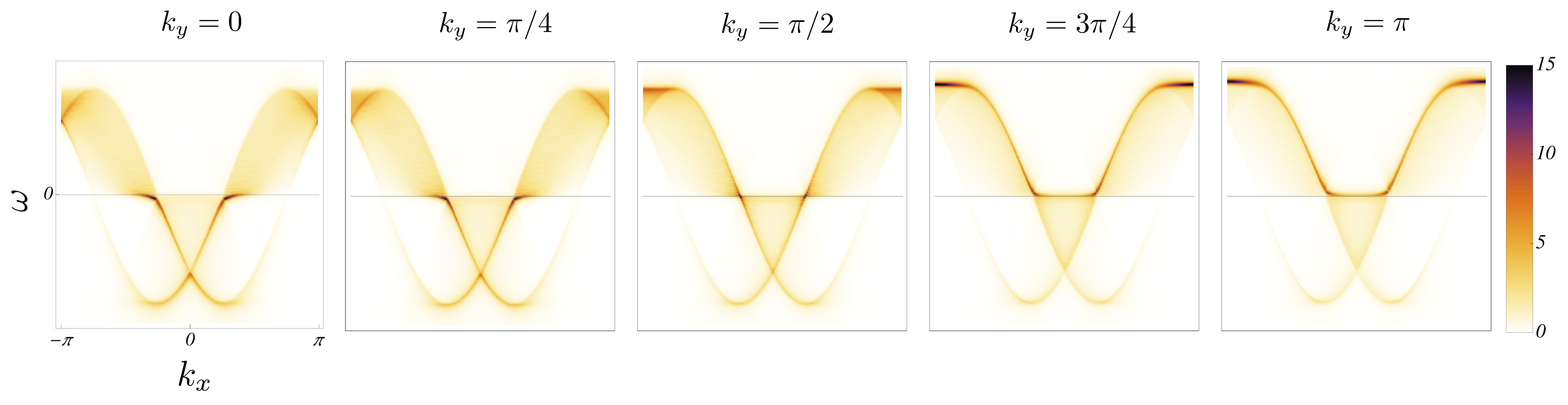} 
\par\end{centering}

\caption{\label{fig:Spectral_HC_np5_Uinf_1} Spectral function of the Hubbard
model for $n=0.5$ and $t'_{1}=0.25$ for the $U=\infty$ where $u_{s}=0$. }

\end{figure*}

At infinite $U$ the spinons are dispersionless ($J_{\text{eff}}=0$)
and the spin velocity vanishes $v_{s}=0$. As a consequence, the lower
edges of the continuum, defined by the spinon dispersion relation,
become flat and the continuum in these regions extends to zero energy.
This is shown in Fig. (\ref{fig:Spectral_HC_np5_Uinf_1}). The central
panel for $k_{y}=\pi/2$ displays, as before, the Hubbard chain spectral
function. Bound states can still form in the regions where the 1D
spectral weight is strictly zero once $t'$ is introduced. However,
this region is smaller than that for the finite $U$ case. Coherent
features also appear at low energies when the bound states enter the
continuum. Considering different values of the transverse momentum
we see the same trends as for finite $U$. In the left panels there
is a \textquotedbl{}refraction\textquotedbl{} of the accumulation
of spectral weight from a \textquotedbl{}spinon\textquotedbl{} branch
line at positive energies (note that it is now a flat line since the
spinon velocity vanishes in the $U\rightarrow\infty$ limit) and a
holon branch at negative energies. In the right panels it is the opposite.
However, the antibound states associated with the holon branch also
sharpen, even though the distribution of the spectral weight through
the continuum is much more visible, as compared to finite $U$. Since
the bound states, associated with the spinons do not concentrate much
spectral weight, this is to be expected.

\section{The role of frustration\label{sec:frustration}}

\begin{figure*}
\begin{centering}
\includegraphics[width=0.9\textwidth]{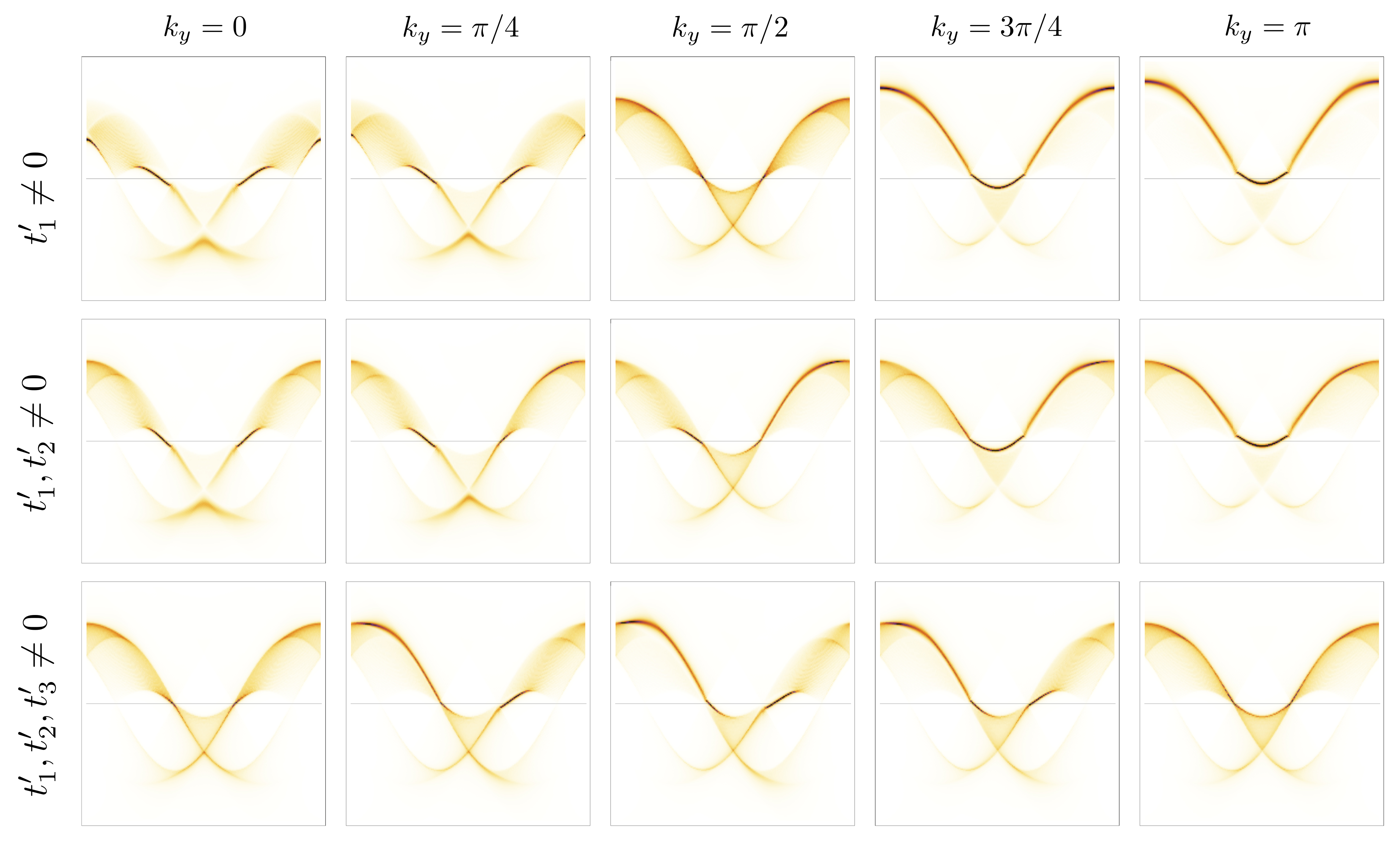} 
\par\end{centering}

\caption{\label{fig:HC_triang_np5_1} Spectral function of the Hubbard model
for several anisotropic lattices: $t'_{1}=0.5,t'_{2}=t'_{3}=0$ (top
row); $t'_{1}=0.25,t'_{2}=0.25,t'_{3}=0$ (central row); $t'_{1}=0.05,t'_{2}=-0.2,t'_{3}=0.25$
(bottom row). Note the decrease of coherent modes and the increase
of continuum-like features as frustration increases. }

\end{figure*}
\begin{figure*}
\begin{centering}
\includegraphics[width=0.9\textwidth]{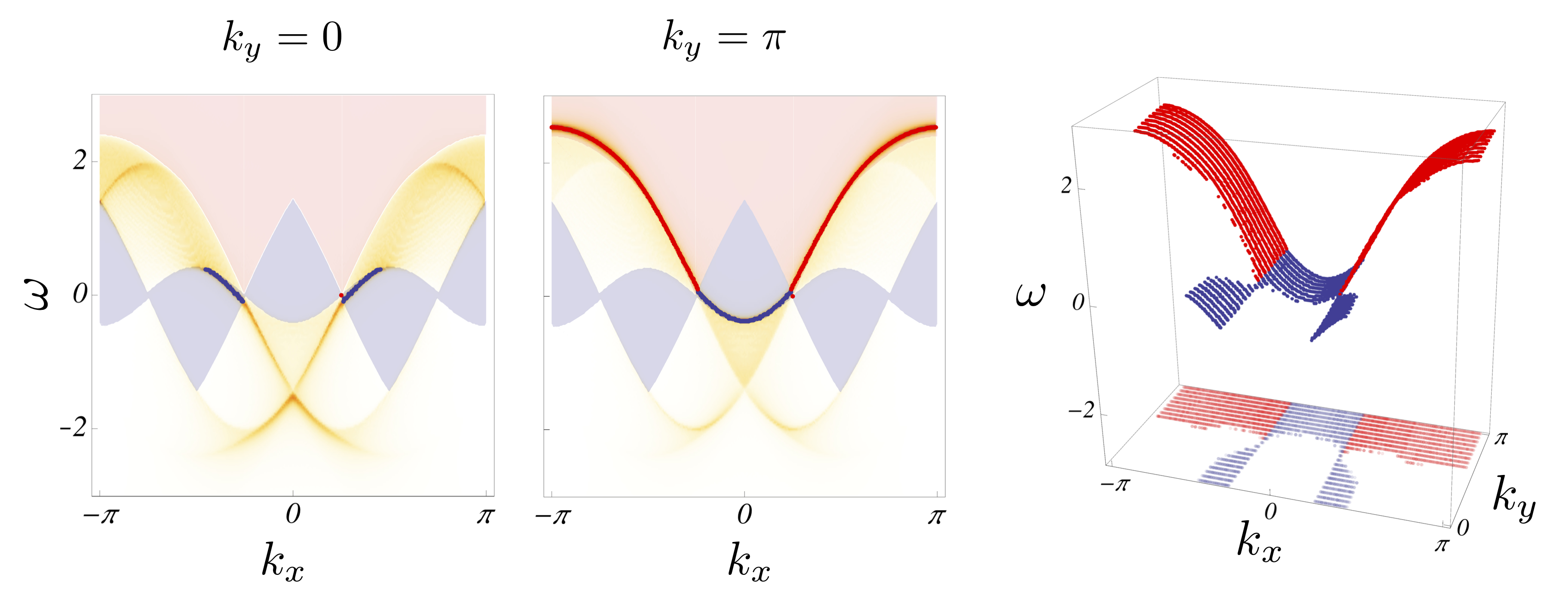} 
\par\end{centering}

\caption{\label{fig:HC_triang_np5_2} Coherent Modes computed for the anisotropic
square lattice with $t'_{1}=0.2$ and $U/t=7.5$. Left Panels: Spectral
function computed for $k_{y}=0$ and $k_{y}=\pi$. Regions with strictly
zero spectral weight are shaded in blue, the boundstate (Blue line)
corresponds to a pole of the 2D Green's function. The high energy
regions, shaded in Red, have low but non-vanishing spectral weight,
therefore the anti-bound states (Red line) arising in this region
have a small but finite width. Right Panel: Dispersion relation of
bound and anti-bound states. }

\end{figure*}

It is interesting to see if frustration, in the sence of addition
of diagonal terms to the rung ladder-like hoppings, has a similar
effect of fractionalization in metallic systems as it does in frustrated
magnetic systems. In this section we investigate the role of frustration
in the finite energy behavior of the system comparing a non-frustrated
lattice (square) with two frustrated lattices, triangular and fully
frustrated.

Fig. \ref{fig:HC_triang_np5_1} shows the spectral function computed
for an anisotropic square $\left(t'_{1}=0.5,t'_{2}=t'_{3}=0\right)$,
triangular $\left(t'_{1}=0.25,t'_{2}=0.25,t'_{3}=0\right)$ and fully
frustrated $\left(t'_{1}=0.05,t'_{2}=-0.2,t'_{3}=0.25\right)$ lattices.
As the number of frustrated links increases, one observes that the
coherent modes are suppressed, as can clearly be seen in Fig. \ref{fig:HC_triang_np5_1}
where the incoherent continuum, typical from the 1D case, carries
much more spectral weight when compared with the anisotropic square
lattice. The reason for the decrease of the coherent features with
the degree of frustration, is easy to understand at RPA level. The
number of bound and anti-bound states due to $t'$, the spectral weight
and the distance of the boundstate from the 1D continuum all grow
with the magnitude of $t'(\mathbf{k})$. Compared to the square lattice,
the values of $t'(\mathbf{k})$ for frustrated systems vary much more
within the Brillouin zone, i.e. even if the maximal value of $\left|t'(\mathbf{k})\right|$
is the same for both lattices, stronger oscillations are expected
for the frustrated case leading to a smaller mean value $\int d\mathbf{k}\left|t'(\mathbf{k})\right|$,
which unfavors the appearance of bound-states.

In order to give a quantitative measure of the coherent modes we computed
the area of the Brillouin zone occupied by the bound and anti-bound
states (see Fig. \ref{fig:HC_triang_np5_2} ). Starting from a square
lattice with $t'_{1}=0.2$ we have increased the total amplitude of
the interchain hopping in three different ways. Table \ref{tab:Area}
shows the evolution of the area of the Brillouin zone covered coherent
modes. For a square lattice, with a larger $t_{1}'$, one observes
a substantial increase of the area occupied by the bound and anti-bound
states: When the same increment is introduced along $t_{2}'$ there
is a small decrease of the area and a substantial decrease is observed
if $t_{3}'$ is further increased.

At low energies, two dimensional spin systems and electronic systems
near half-filling (where they can be well described by $t-J$ like
models), are expected to be rather sensitive to frustration and may
develop exotic spin-liquid phases with non-FL behavior. Even if we
do not study this low energy regimes, the results presented here do
point out that the finite energy spectrum is significantly affected
by the frustrated nature of the lattice even if the interchain hopping
is small compared with the monitored energy scale.

\begin{table}
\begin{centering}
\begin{tabular}{|c|c|c|}
\cline{2-2} 
\multicolumn{1}{c|||}{} & $t'_{1}=0.2$  & \multicolumn{1}{c}{}\tabularnewline
\multicolumn{1}{c|||}{} & $t'_{2}=0.0$  & \multicolumn{1}{c}{}\tabularnewline
\multicolumn{1}{c|||}{} & $t'_{3}=0.0$  & \multicolumn{1}{c}{}\tabularnewline
\hline 
$t'_{1}=0.3$  & $t'_{1}=0.2$  & $t'_{1}=0.2$\tabularnewline
$t'_{2}=0.0$  & $t'_{2}=0.1$  & $t'_{2}=0.1$\tabularnewline
$t'_{3}=0.0$  & $t'_{3}=0.0$  & $t'_{3}=0.1$\tabularnewline
$\mathbf{\left(+11.0\%\right)}$  & $\mathbf{\left(-0.07\%\right)}$  & $\mathbf{\left(-16.0\%\right)}$\tabularnewline
\hline
\end{tabular}
\par\end{centering}

\caption{\label{tab:Area}Evolution of the area of the Brillouin zone covered
by coherent modes for different values of interchain hopping. The
percentage values are relative to the area of the square lattice with
$t'_{1}=0.2,t'_{2}=t'_{3}=0.0$ . }

\end{table}

\section{Discussion\label{sec:Discussion}}

The unusual non-Fermi liquid like properties of some two-dimensional
strongly correlated systems has lead to the proposal that some signatures
of the exotic properties of one-dimensional systems may be observed
in their two-dimensional counterparts. The dimensional crossover from
one to two dimensions has been considered by several authors and in
most cases it has been found that the one-dimensional features are
to a large degree lost, particularly at low energies. One characteristic
of the one-dimensional systems is the fractionalization of degrees
of freedom which has, however, been shown to persist in some frustrated
magnetic systems via the deconfinement of spinons, instead of the
coherent magnon-like degrees of freedom characteristic of higher dimensional
systems. This apparent fractionalization has been confirmed recently
as shown, for instance, in \cite{Kohno_2007}.

In this work we have considered Hubbard chains coupled in non-frustrated
and frustrated ways and have studied the quasiparticle properties
via the spectral function. In order to study the crossover from one
to two dimensions we considered spatially anisotropic systems where
the interchain couplings (hoppings) are small compared to the intrachain
hoppings. The spectral function of the one-dimensional Hubbard model
is in general hard to solve but, in some limits, it can be obtained
exactly/approximatively such as in the infinite/large U limits. This
solution was used to obtain, in the RPA, the two-dimensional spectral
function. In the low energy regime a small interchain hopping leads
to the formation of a Fermi surface, as shown before by other authors.
The appearance of bound states leads to a significant concentration
of spectral weight, that extends in some cases to finite energies
in a way similar to the formation of coherent modes, as expected in
a Fermi liquid like system. However a significant weight is also observed
spread through a continuum characteristic of fractionalization of
degrees of freedom. This is particularly found when there is frustration
in the hoppings, as evidenced by the increase in spectral weight out
of the bound states as frustration increases.

It would be interesting to compare these results with experimental
results for anisotropic conductors. However, to our knowledge, such
systems have not been identified. Some systems show anisotropy but
they are not weakly coupled, such as the BEDT systems \cite{Kandpal}.
We expect however that with the advent of fermionic cold atoms in
optical lattices the predictions of this work may be tested and new
classes of exotic two-dimensional systems may be found.
\begin{acknowledgments}
We acknowledge discussions with Shi-Jian Gu at an early stage of this
work, discussions with J. Carmelo, a discussion with Collin Broholm
and the hospitality of Hai-Qing Lin and Shi-Jian Gu. This research
was partially supported by the ESF Program INSTANS, by FCT-Portugal
through grants PTDC/FIS/64926/2006 and PTDC/FIS/70843/2006, and by
the Hungarian OTKA Grant No. K73455. PR acknowledges support through
FCT BPD grant SFRH/BPD/43400/2008. 
\end{acknowledgments}
\appendix
%dummy comment inserted by tex2lyx to ensure that this paragraph is not empty

\section{Spectral function in restricted Hilbert space \label{sec:Ap_restricted}}

In this section we discuss the extension of a method, introduced in
\cite{Kohno_2007} to study anisotropic anti-ferromagnets, to the
case of electronic systems. This method permits to write down the
2D spectral function as a function of 1D quantities, in the limit
of small inter-chain coupling. The main ingredient is to restrict
the Hilbert space to the subspace spanned by eigenstates of decoupled
chains with few spinon-chargon pairs. In doing so one neglects some
processes that are expected to carry low spectral weight. Besides
their formal final resemblance, the expression for the 2D spectral
function obtained in this way, follows from fundamentally different
approximations than the ones leading to the RPA result. However, we
will show explicitly that some problems arise when dealing with this
approach, that lead to inconsistencies that prevented us from applying
this method.

From the exact one dimensional solution of the Hubbard model one finds
a multitude of excitations that can be identified as coming from charge
and spin degrees of freedom. However, for practical purposes, single
spin-charge excitation characterized by their rapidities carry the
vast majority ($>95\%$) of the spectral weight (see \cite{Carmelo_2006-1}).
From small to moderate inter-chain coupling, if no phase transition
occurs, such excitations are expected to preserve their identity furnishing
a natural basis for perturbation theory. The physical picture of the
perturbed excitations is given by one dimensional fractionalized electron
(or hole) that hops coherently between neighboring chains. These two
facts: small inter-chain coupling and low spectral weight of the other
types of excitations, allow significant simplification of the problem.
The former allows an expansion in small inter-chain coupling and the
latter justifies the truncation of the Hilbert space to two particle
states.

\subsection{Two-particle states}

Let the ground state of the unperturbed system ($t'_{i}=0$) be denoted
by $\ket{\mathbf{0}}=\otimes_{y}\ket{0,y}$, with $\ket{0,y}$ the
ground state of the Hubbard Hamiltonian for chain $y$. Its energy
is $E_{0}L_{x}L_{y}$, with $E_{0}$ the mean energy per site and
$L_{x}$ and $L{}_{y}$ the number of sites in the $x$ and $y$ directions,
respectively. From the Bethe Ansatz solution, the two particle states
are labeled by the charge and spin rapidities $\nu^{(c)},\nu^{(s)}$
, by the value of the $S_{z}$ component of the spin $\sigma$ and
by the total charge of the state $b=\pm1$, compared to the GS. Such
states can be alternatively labeled by their total energy and momentum
$\ket{\varepsilon_{l},k_{x},\sigma}=\ket{\nu^{(c)},\nu^{(s)},\sigma}$
where $\varepsilon_{l}>0$ and $k_{x}$ are defined by \begin{align}
H_{1D}\ket{\varepsilon_{l},k_{x},\sigma,b} & =\left(\varepsilon_{l}+E_{0}L_{x}\right)\ket{\varepsilon_{l},k_{x},\sigma,b}\\
T\ket{\varepsilon_{l},k_{x},\sigma,b} & =e^{ik_{x}}\ket{\varepsilon_{l},k_{x},\sigma,b}\end{align}
 where $T$ is the operator that translates the system by one lattice
site. For sake of clarity a finite system is considered at this stage,
the thermodynamic limit being taken only in the final results; therefore,
$\varepsilon_{l}$ and $k_{x}$ are taken within a discrete set of
values. The two particle states of the 2D system with momentum $\mathbf{k}=\left(k_{x},k_{y}\right)$
are defined as the Fourier transform in the $y$ direction of the
states with only one excited chain: \begin{equation}
\ket{\varepsilon_{l},\mathbf{k},\sigma,b}=\frac{1}{\sqrt{L_{y}}}\sum_{y}e^{ik_{y}y}\ket{\varepsilon_{l},k_{x},\sigma,b;y}\otimes_{y'\neq y}\ket{0,y'}\end{equation}
 By construction these states are orthogonal to each other as well
as to the unperturbed GS, $\ket{\mathbf{0}}$. The projector to the
subspace spanned by the two-particle states and the $t'=0$ GS is
denoted $\mathbb{P}_{0+2}=\ket{\mathbf{0}}\bra{\mathbf{0}}+\sum_{\mathbf{k},\sigma,l}\ket{\varepsilon_{l},\mathbf{k},\sigma}\bra{\varepsilon_{l},\mathbf{k},\sigma}$.

\subsection{Spectral Function}

The 2D spectral function is obtained as the imaginary part of the
retarded Green's function $Sp_{\sigma}(\omega,\mathbf{k})=-\frac{1}{\pi}\lim_{\eta\to0}\im G_{\sigma}^{R}(\omega+i\eta,\mathbf{k})$
defined as \begin{eqnarray}
 &  & G_{\sigma}^{R}(\omega+i\eta,\mathbf{q})\nonumber \\
 & = & \int_{0}^{\infty}dt\ e^{i\left(\omega+i\eta\right)t}(-i)\bra{\tilde{\mathbf{0}}}c_{\mathbf{q},\sigma}(t)c_{\mathbf{q},\sigma}^{\dagger}(0)+c_{\mathbf{q},\sigma}^{\dagger}(0)c_{\mathbf{q},\sigma}(t)\ket{\tilde{\mathbf{0}}}\nonumber \\
 & = & \sum_{n,\mathbf{k},\sigma',b}\left[\frac{\left|\bra{\tilde{\mathbf{0}}}c_{\mathbf{q},\sigma}\ket{\Psi_{\mathbf{k},\sigma,b}^{(n)}}\right|^{2}}{\omega-\delta E_{\mathbf{k},\sigma,b}^{(n)}+i\eta}+\frac{\left|\bra{\tilde{\mathbf{0}}}c_{\mathbf{q},\sigma}^{\dagger}\ket{\Psi_{\mathbf{k},\sigma,b}^{(n)}}\right|^{2}}{\omega+\delta E_{\mathbf{k},\sigma,b}^{(n)}+i\eta}\right]\nonumber \\
\label{eq:GF_1}\end{eqnarray}
 with $\ket{\tilde{\mathbf{0}}}$ the exact GS of the coupled chains.
The second equality was obtained using a complete set of eigenstates
$\ket{\Psi_{\mathbf{k},\sigma,b}^{(n)}}$ with energy $E_{\mathbf{k},\sigma,b}^{(n)}=\delta E_{\mathbf{k},\sigma,b}^{(n)}+L_{x}L_{y}E_{0}$.
Both $\ket{\tilde{\mathbf{0}}}$ and $\ket{\Psi_{\mathbf{k},\sigma,b}^{(n)}}$
will be approximated by their projection in the considered subspace
and the effective Hamiltonian is given by $H_{\text{eff}}=\mathbb{P}_{0+2}H\mathbb{P}_{0+2}$.

Using first order perturbation theory in the two-particle subspace
one finds: \begin{align}
\ket{\tilde{\mathbf{0}}} & \simeq\ket{\mathbf{0}}+\frac{1}{E_{0}-H_{\parallel}}\mathbb{P}_{0+2}H_{\perp}\mathbb{P}_{0+2}\ket{\mathbf{0}}+...=\ket{\mathbf{0}}+O\left(t'^{2}\right)\label{eq:GS_corr}\end{align}
 where the last equality follows since $H_{\perp}$ acting on $\ket{\mathbf{0}}$
creates two electron-like excitations in neighboring chains which
are out of the subspace. Therefore no corrections to the decoupled
GS arise in first order in $t'$ within the considered subspace. Since
$H_{\text{eff}}$ does not couple states with different momentum,
total spin or charge one can decompose the eigenstates as \begin{equation}
\ket{\Psi_{\mathbf{k},\sigma,b}^{(n)}}=\sum_{l}\ \psi_{\mathbf{k},\sigma,b}(\varepsilon_{l})\ket{\mathbf{k},\varepsilon_{l},\sigma,b}.\end{equation}
 where the summation index runs only over the 1D eigen energies. Computing
the matrix elements of $H_{\text{eff}}$ one finds the Schrödinger
equation for the amplitudes

\begin{eqnarray}
 &  & \psi_{\mathbf{k},\sigma,b}(\varepsilon_{l})\left(\varepsilon_{l}-\delta E_{\mathbf{k},\sigma,b}^{(n)}\right)+b\, t'(\mathbf{k})\bar{A}_{b,\sigma}\left(\varepsilon_{l},k_{x}\right)\sum_{l'}\ A_{b,\sigma}\left(\varepsilon_{l'},k_{x}\right)\psi_{\mathbf{k},\sigma,b}(\varepsilon_{l'})=0\label{eq:schroedinger}\end{eqnarray}
 where \begin{eqnarray}
A_{b=-1,\sigma}\left(\varepsilon_{l},k_{x}\right) & = & \bra 0c_{k_{x},\sigma}\ket{k_{x},\varepsilon_{l},\sigma,b}\label{eq:A's_0}\\
A_{b=1,\sigma}\left(\varepsilon,k_{x}\right) & = & \bra 0c_{k_{x},\sigma}^{\dagger}\ket{k_{x},\varepsilon_{l},\sigma,b}\label{eq:A's}\end{eqnarray}
 are pure one dimensional quantities and $t'(\mathbf{k})$ is the
Fourier transform of the transverse hopping matrix (\ref{eq:t_k}).
For completeness the 1D Green's function in this notation is given
by\begin{eqnarray*}
G_{1D,\sigma}^{R}\left(\omega+i\eta,k_{x}\right) & = & \sum_{b=\pm}\sum_{l}\frac{A_{b,\sigma}\left(\varepsilon_{l},k_{x}\right)\bar{A}_{b,\sigma}\left(\varepsilon_{l},k_{x}\right)}{\omega+b\,\varepsilon_{l}+i\eta}.\end{eqnarray*}
 Defining $B_{\mathbf{k},\sigma,b}^{(n)}=\sum_{l}\ A_{b,\sigma}\left(\varepsilon_{l},k_{x}\right)\psi_{\mathbf{k},\sigma,b}(\varepsilon_{l})$
and using Eqs. (\ref{eq:GF_1},\ref{eq:A's_0},\ref{eq:A's}) the
approximate 2D Green's function can now be written: \begin{eqnarray}
G_{\sigma}^{R}(\omega,\mathbf{k}) & = & \sum_{n,b}\frac{B_{\mathbf{k},\sigma,b}^{(n)}\bar{B}_{\mathbf{k},\sigma,b}^{(n)}}{\omega+b\,\delta E_{\mathbf{k},\sigma,b}^{(n)}+i\eta},\label{eq:green_1}\end{eqnarray}
 which coincides with the 1D case when $t'=0$. Moreover, the particular
form of Eq. (\ref{eq:schroedinger}) allows the derivation of the
following identities: \begin{eqnarray}
1 & = & \, t'(\mathbf{k})\sum_{l}\frac{A_{b,\sigma}\left(\varepsilon_{l},k_{x}\right)\bar{A}_{b,\sigma}\left(\varepsilon_{l},k_{x}\right)}{b\left(\delta E_{\mathbf{k},\sigma,b}^{(n)}-\varepsilon_{l}\right)},\\
\nonumber \\\left[t'(\mathbf{k})^{2}\, B_{\mathbf{k},\sigma,b}^{(n)}\bar{B}_{\mathbf{k},\sigma,b}^{(n)}\right]^{-1} & = & \sum_{l}\frac{A_{b,\sigma}\left(\varepsilon_{l},k_{x}\right)\bar{A}_{b,\sigma}\left(\varepsilon_{l},k_{x}\right)}{\left(\delta E_{\mathbf{k},\sigma,b}^{(n)}-\varepsilon_{l}\right)^{2}},\end{eqnarray}
 where the first equality is obtained by simple manipulations of Eq.
(\ref{eq:schroedinger}) and the second follows from imposing unit
norm to the eigenstates. These equalities enable the definition of
the complex valued functions\begin{eqnarray*}
F_{\mathbf{k},\sigma,b}(z) & = & \sum_{l}\frac{A_{b,\sigma}\left(\varepsilon_{l},k_{x}\right)\bar{A}_{b,\sigma}\left(\varepsilon_{l},k_{x}\right)}{z-b\,\varepsilon_{l}},\end{eqnarray*}
 with the properties

\begin{eqnarray}
F_{\mathbf{k},\sigma,b}(b\,\delta E_{\mathbf{k},\sigma,b}^{(n)}) & = & \left[t'(\mathbf{k})\right]^{-1},\\
F_{\mathbf{k},\sigma,b}'(b\,\delta E_{\mathbf{k},\sigma,b}^{(n)}) & = & -\left[t'(\mathbf{k})^{2}\, B_{\mathbf{k},\sigma,b}^{(n)}\bar{B}_{\mathbf{k},\sigma,b}^{(n)}\right]^{-1}.\end{eqnarray}
 So that for a test function $\rho(z)$, analytic in the vicinity
of the real line, one has

\begin{eqnarray}
 &  & \frac{1}{2\pi i}\oint\ dz\ \rho(z)\,\frac{1}{\left[F_{\mathbf{k},\sigma,b}(z)\right]^{-1}-\, t'(\mathbf{k})}=\\
 & = & \sum_{n}\rho(b\,\delta E_{\mathbf{k},\sigma,b}^{(n)})\, B_{\mathbf{k},\sigma,b}^{(n)}\bar{B}_{\mathbf{k},\sigma,b}^{(n)}\end{eqnarray}
 where the contour is taken in the domain of analyticity of $\rho(z)$
and encircles anti-clockwise all eigen energies $b\,\delta E_{\mathbf{k},\sigma,b}^{(n)}$.
In particular, using $\rho(z)=\frac{1}{\omega-z+i\eta}$, the Green's
function (\ref{eq:green_1}) can be written as \begin{eqnarray}
 &  & G_{\sigma}^{R}(\omega+i\eta,\mathbf{k})=\label{eq:green_contour}\\
 &  & \frac{1}{2\pi i}\sum_{b}\oint dz\ \frac{1}{\left[F_{\mathbf{k},\sigma,b}(z)\right]^{-1}-\, t'(\mathbf{k})}\frac{1}{\omega-z+i\eta}=\nonumber \\
 &  & =\sum_{b}\frac{1}{\left[F_{\mathbf{k},\sigma,b}(\omega+i\eta)\right]^{-1}-\, t'(\mathbf{k})}\nonumber \end{eqnarray}
 where the contour does not including the $\omega+i\eta$ pole.

A remark about this method is in order at this point. Note the RPA
expression given by Eq. (\ref{eq:RPA}) so the differences between
the two approaches can be clearly observed. Contrarily to the RPA
it is not possible to define a single analytic function $F$ gathering
both positive and negative energy contributions. This derives from
the fact that in the present method Eq.(\ref{eq:green_contour}) cannot
be given as a function of the 1D Green's function. Instead, each branch
has to be summed separately in Eq. (\ref{eq:green_contour}) in order
to obtain the same result as in Eq. (\ref{eq:green_1}), which is
itself a consequence of the fact that both $b=\pm$ sectors are uncoupled
by the Schrödinger equation. Care must be taken when the bound states
cross $\omega=0$ in (\ref{eq:green_contour}); this would correspond
to negative excitation energies arising in the Schrödinger equation,
signaling an instability of the system. Even though it is still possible
to give an expression for the Green's function in this case, it would
not be physically justified to use this result. Since this happens
somewhere in the Brillouin zone for the Hubbard model it prevented
us to use this method to compute the 2D spectral function. For further
comparison we give the spectral and the Green's function computed
with both methods as a function of the 1D spectral function:

\begin{tabular}{|c|c|}
\hline 
Restricted Subspace  & RPA\tabularnewline
\hline 
$\begin{array}{c}
\\G_{\sigma}^{R}(\omega+i\eta,\mathbf{k})=\sum_{b=\pm1}\frac{1}{\left[F_{\mathbf{k},\sigma,b}(\omega+i\eta)\right]^{-1}-\, t'(\mathbf{k})}\\
\\\text{where}\\
F_{\mathbf{k},\sigma,b}(\omega+i\eta)=\int d\nu\,\frac{Sp_{1D}\left(\nu,k_{x}\right)\theta\left(b\nu\right)}{\omega+i\eta-\nu}\\
\\\end{array}$  & $\begin{array}{c}
\\G_{\sigma}^{R}(\omega+i\eta,\mathbf{k})=\frac{1}{\left[G_{1D}(\omega+i\eta,k_{x})\right]^{-1}-\, t'(\mathbf{k})}\\
\\\text{where}\\
G_{1D}(\omega+i\eta,k_{x})=\int d\nu\,\frac{Sp_{1D}\left(\nu,k_{x}\right)}{\omega+i\eta-\nu}\\
\\\end{array}$\tabularnewline
\hline 
$\begin{array}{c}
\\Sp(\varepsilon,\mathbf{k})=\sum_{b=\pm1}\frac{\chi_{\mathbf{k},\sigma,b}^{''}(\varepsilon)/\pi}{\left[1-t'(\mathbf{k})\chi_{\mathbf{k},\sigma,b}^{'}(\varepsilon)\right]^{2}+\left[t'(\mathbf{k})\chi_{\mathbf{k},\sigma,b}^{''}(\varepsilon)\right]^{2}}\\
+\sum_{p,b}\frac{1}{F_{\mathbf{k},\sigma,b}^{-1'}(z_{p})}\delta\left(\varepsilon-z_{p}\right)\\
\\\text{where}\\
\chi_{\mathbf{k},\sigma,b}^{''}(\varepsilon)=\pi Sp_{1D}\left(\varepsilon,k_{x}\right)\theta\left(b\varepsilon\right)\\
\chi_{\mathbf{k},\sigma,b}^{'}(\varepsilon)=P\int d\nu\,\frac{Sp_{1D}\left(\nu,k_{x}\right)\theta\left(b\varepsilon\right)}{\varepsilon-\nu}\\
\\\end{array}$  & $\begin{array}{c}
\\Sp(\varepsilon,\mathbf{k})=\frac{\chi_{\mathbf{k},\sigma}^{\left(\text{RPA}\right)''}(\varepsilon)/\pi}{\left[1-t'(\mathbf{k})\chi_{\mathbf{k},\sigma}^{\left(\text{RPA}\right)'}(\varepsilon)\right]^{2}+\left[t'(\mathbf{k})\chi_{\mathbf{k},\sigma}^{\left(\text{RPA}\right)''}(\varepsilon)\right]^{2}}\\
+\sum_{p}\frac{1}{G_{1D}^{-1'}(z_{p},k_{x})}\delta\left(\varepsilon-z_{p}\right)\\
\\\text{where}\\
\chi_{\mathbf{k},\sigma}^{\left(\text{RPA}\right)''}(\varepsilon)=\pi Sp_{1D}\left(\varepsilon,k_{x}\right)\\
\chi_{\mathbf{k},\sigma}^{\left(\text{RPA}\right)'}(\varepsilon)=P\int d\nu\,\frac{Sp_{1D}\left(\nu,k_{x}\right)}{\varepsilon-\nu}\\
\\\end{array}$\tabularnewline
\hline
\end{tabular}

In the above expressions the thermodynamic limit was taken replacing
$\sum_{l}$$\rho(\varepsilon_{l},k_{x},\sigma,b)$ by $\int d\nu\ D_{k_{x},\sigma,b}\left(\nu\right)\rho(\nu)$,
where $D_{k_{x},\sigma,b}\left(\nu\right)$ is the one dimensional
density of states with quantum numbers $k_{x},\sigma,b$, and using
the definition $Sp_{1D}\left(\varepsilon,k_{x}\right)=\sum_{b}D_{k_{x},\sigma,b}\left(b\varepsilon\right)A_{b,\sigma}\left(b\varepsilon,k_{x}\right)\bar{A}_{b,\sigma}\left(b\varepsilon,k_{x}\right)$.
Note that when this replacement is done, the Green's function acquires
a branch cut in the support of $D_{k_{x},\sigma,b}\left(b\nu\right)$
and coherent contributions from the simple poles for both methods.
Corrections to the effective Hamiltonian method can be included as
in Ref. in \cite{Kohno_2007} by considering a larger subspace spanned
by states containing higher ordered processes along a chain and/or
where more than one chain is in an excited state. In the generic case
the GS will also have corrections (see Eq. (\ref{eq:GS_corr})) of
higher order in $t'$.

\section{Factorized wave function and spectral function\label{sec:Ap_Factorized}\label{Sec:Appendix:factorizedWF}}

In the $U\rightarrow+\infty$ limit of the Hubbard model the doubly
occupied sites are forbidden, and the electrons with opposite spins
cannot jump over each other - the sequence of the spins of the electrons
is fixed. As a consequence, the wave functions can be written in a
factorized form, \begin{equation}
|\Psi_{P}^{N}\rangle=|\psi_{L,Q}^{N}(\{I\})\rangle\otimes|\chi_{N}^{N_{\downarrow}}(Q,\tilde{f}_{Q})\rangle\end{equation}
 where $|\chi\rangle$ stands for the spin-part of the $N$ electrons
that is defined on a fictitious lattice of the $N$ sites, with the
wave vector $Q=2\pi K/N$ ($K=0,\dots,N-1$ is an integer) and $\tilde{f}_{Q}$
are some other quantum numbers \cite{Woynarovich82,Ogata_1990}. The
$|\psi\rangle$ describes the $N$ electrons as spinless free fermions
with twisted boundary condition imposed by the spins: \begin{equation}
Lk_{j}=2\pi I_{j}+Q\end{equation}
 where the wave vector $Q$ of the spin wave function appears as a
phase shift and $I_{j}=0,\dots,L-1$. The total momentum and energy
of the state are given by \begin{eqnarray*}
P & = & \sum_{j}k_{j}=\frac{2\pi}{L}\sum_{j}I_{j}+\frac{N}{L}Q=\frac{2\pi}{L}\left(\sum I_{j}+K\right)\\
E & = & -2t\sum_{j}\cos k_{j}\end{eqnarray*}
 Strictly speaking, for $U=+\infty$ all the spin wave function are
degenerate in energy, and the $U\rightarrow+\infty$ limit is taken
such that in the ground state the $|\chi\rangle$ coincides with the
ground state of the Heisenberg model, with wave vector $Q=\pi$.

The electron addition and removal spectral functions are then given
as \begin{eqnarray}
A(k,\omega) & = & \sum_{Q}C(Q)A_{Q}(k,\omega)\\
B(k,\omega) & = & \sum_{Q}D(Q)B_{Q}(k,\omega)\end{eqnarray}
 where the $A_{Q}(k,\omega)$ and $B_{Q}(k,\omega)$ are coming from
the charge, and $C(Q)$ and $D(Q)$ from the spin part of the wave
function and can be evaluated as described in Ref.~\cite{Penc_1997}.
We note the absence of the energy scale in the spin part.

For finite values of $U/t$ the spin part gets finite dispersion.
As it has been noted in \cite{Penc_1997} (see also Ref. \cite{talstra}),
in the $\omega$-resolved $C(Q,\omega)$ and $D(Q,\omega)$ the weight
is to large extent concentrated on the lower edge of the continuum,
following the dispersion of the one-spinon branch (we note that the
number of spinons in the final states is odd, since we had added one
spinon to the initial spin wave function), so that \begin{eqnarray*}
C(Q,\omega) & = & C(Q)\delta(\omega-\varepsilon_{s}-\varepsilon_{Q}),\\
D(Q,\omega) & = & D(Q)\delta(\omega-\varepsilon_{s}+\varepsilon_{Q}),\end{eqnarray*}
 where $\varepsilon_{Q}$ is the des Cloizeaux-Pearson dispersion\cite{desCLoizeaux,spinon}
\begin{equation}
\varepsilon_{Q}=\frac{\pi}{2}J_{\text{eff}}|\sin(Q-\pi/2)|\;,\label{eq:dCP}\end{equation}
 and \begin{equation}
J_{\text{eff}}=n\left(1-\frac{\sin2\pi n}{2\pi n}\right)\frac{4t^{2}}{U}\end{equation}
 is the effective exchange in the $N$-site Heisenberg model of the
spin part. After the convolution with the $A_{Q}(\omega,k)$ and $B_{Q}(\omega,k)$
charge part, the inclusions of the spinon dispersion given above provides
a finite spinon dispersion that is seen in Fig. \ref{fig:Spectral_HC_np5_1}:
it defines the lower edge of the $A(\omega,k)$ for the $k$ values
between the $k_{F}$ and the $3k_{F}$, and the lower edge of the
$B(\omega,k)$ for $-k_{F}<k<k_{F}$.

\section{RPA and next to leading order corrections\label{sec:Ap_RPA}}

In this section we rederive the RPA results obtained before by many
authors and give explicitly the next to leading order corrections.
However, since 1D correlation functions of higher order are needed
in order compute these corrections they were not included in the computation
of the spectral function in the main text.

The partition function of the model with Grassmanian sources is given
by %\begin{widetext} 
\begin{eqnarray}
Z\left[\zeta^{\dagger},\zeta\right] & = & \int Dc^{\dagger}Dc\ e^{-\int_{\tau}\left[\sum_{\mathbf{k},\sigma}c_{\mathbf{k},\sigma}^{\dagger}(\tau)\partial_{\tau}c_{\mathbf{k},\sigma}(\tau)+:H(\tau):-\sum_{\mathbf{k},\sigma}\left(\zeta_{\mathbf{k},\sigma}^{\dagger}(\tau)c_{\mathbf{k},\sigma}(\tau)+c_{\mathbf{k},\sigma}^{\dagger}(\tau)\zeta_{\mathbf{k},\sigma}(\tau)\right)\right]}\nonumber \\
 & = & Z_{\parallel}\av{e^{\mathbf{c}^{\dagger}.\mathbf{t}'.\mathbf{c}+\boldsymbol{\zeta}^{\dagger}.\mathbf{c}+\mathbf{c}^{\dagger}.\boldsymbol{\zeta}}}_{\parallel}\end{eqnarray}
 %\end{widetext}
where $Z_{\parallel}$ and $\av{...}_{\parallel}$ are respectively
the partition function and the expectation value of an operator in
absence of interchain coupling and $H=H_{\parallel}+\sum_{ij}c_{i}^{\dagger}t'_{ij}c_{j}$.
The compact notation $\mathbf{c}^{\dagger}.\mathbf{t}'.\mathbf{c}=\int d\tau\ \sum_{\mathbf{k},\sigma}c_{\mathbf{k},\sigma}^{\dagger}(\tau)t'\left(\mathbf{k}\right)c_{\mathbf{k},\sigma}(\tau)$
and $\boldsymbol{\zeta}^{\dagger}.\mathbf{c}=\int d\tau\sum_{\mathbf{k},\sigma}\zeta_{\mathbf{k},\sigma}^{\dagger}(\tau)c_{\mathbf{k},\sigma}(\tau)$
$ $ was introduced to improve the readability of the expressions
and will be used in the rest of this section. Inserting a Grassmanian
Hubbard-Stratonovich (HS) field $\psi$ to decouple the hopping term
and performing a subsequent shift in this field in order to let the
term within brackets independent from the sources one gets \begin{align}
Z\left[\zeta^{\dagger},\zeta\right] & =Z_{\parallel}\int D\psi^{\dagger}D\psi\ e^{-N\, F}\label{eq:Z_sp}\end{align}
 with \begin{eqnarray}
F & = & -\left(\boldsymbol{\psi}^{\dagger}+\boldsymbol{\zeta}^{\dagger}\right).\mathbf{t}'^{-1}.\left(\boldsymbol{\psi}+\boldsymbol{\zeta}\right)-\ln\av{e^{\boldsymbol{\psi}^{\dagger}\mathbf{c}+\mathbf{c}^{\dagger}\boldsymbol{\psi}}}_{\parallel}.\nonumber \\
\end{eqnarray}
 $N=1$ for the physical case, but we will nevertheless perform a
saddle point approximation in Eq. (\ref{eq:Z_sp}) which can be seen
as an expansion around $N\to\infty$. This method is similar to the
one considered in \cite{Boies_1995}. When the HS variables are bosons
such procedure is equivalent to a given mean field decoupling. The
saddle point value is defined by $\delta_{\boldsymbol{\psi}}F=\left(\boldsymbol{\psi}^{\dagger}+\boldsymbol{\zeta}^{\dagger}\right).\mathbf{t}'^{-1}+\av{\mathbf{c}^{\dagger}}_{\parallel\psi}$
with $\av{...}_{\parallel\psi}=\av{e^{\boldsymbol{\psi}^{\dagger}\mathbf{c}+\mathbf{c}^{\dagger}\boldsymbol{\psi}}}_{\parallel}^{-1}\av{...\, e^{\boldsymbol{\psi}^{\dagger}\mathbf{c}+\mathbf{c}^{\dagger}\boldsymbol{\psi}}}_{\parallel}$.
Assuming that $t'\ll1$ the saddle point value is $\psi=0$ when computed
at $\zeta=0$. To quadratic order one obtains \begin{eqnarray}
Z\left[\zeta^{\dagger},\zeta\right] & = & Z_{\parallel}e^{-N\,\left\{ F-\frac{1}{N}\frac{1}{2}\tr\ln\left[-\delta^{2}F\right]\right\} }\times\nonumber \\
 &  & \times\left[1+O\left(\frac{1}{N}\right)\right]\end{eqnarray}
 where $\delta^{2}F_{i,j}=\delta_{\Psi_{i}}\delta_{\Psi_{j}}F$ with
$\Psi=(\psi,\psi^{\dagger})$ is the second derivative matrix \begin{eqnarray*}
\delta^{2}F_{\bar{i},j} & = & \delta_{\Psi_{i}^{\dagger}}\delta_{\Psi_{j}}F\\
 & = & \left(\begin{array}{cc}
\mathbf{t}'^{-1}+\av{\mathbf{c}\mathbf{c}^{\dagger}}_{\parallel} & -\av{\mathbf{c}\mathbf{c}}_{\parallel}\\
\av{\mathbf{c}^{\dagger}\mathbf{c}^{\dagger}}_{\parallel} & -\mathbf{t}'^{-1}+\av{\mathbf{c}^{\dagger}\mathbf{c}}_{\parallel}\end{array}\right)_{i,j}\end{eqnarray*}
 which is diagonal since the anomalous terms vanish. The Green's function
is obtained taking derivatives with respect to the sources \begin{eqnarray*}
G_{\alpha,\alpha'} & = & \frac{1}{N}\left.d_{\zeta_{\alpha}^{\dagger}}d_{\zeta_{\alpha'}}\ln Z\left[\zeta^{\dagger},\zeta\right]\right|_{\zeta=0}\\
 & = & \left.-\delta_{\zeta_{\alpha}^{\dagger}}\delta_{\zeta_{\alpha'}}F+\frac{1}{2N}\tr\left[\delta^{2}F^{-1}\left(\delta_{\zeta_{\alpha}^{\dagger}}\delta_{\zeta_{\alpha'}}\delta^{2}F\right)\right]\right|_{\zeta=0}\\
 &  & +O\left(\frac{1}{N^{2}}\right)\end{eqnarray*}
 where $d_{\zeta_{\alpha}^{\dagger}}=\delta_{\zeta_{\alpha}^{\dagger}}+\left(\delta_{\zeta_{\alpha}^{\dagger}}\Psi_{i}\right)\delta_{\Psi_{i}}$
stands for the total variation and $\delta$ for explicit one. Using
the saddle point condition $\delta_{\Psi_{j}}F=0$ and total variations
of this relation one obtains\begin{eqnarray}
G_{\alpha,\alpha'}^{-1} & = & \tilde{G}_{\alpha\alpha'}^{-1}-\frac{1}{N}\tilde{G}_{m'n'}\tilde{\Gamma}_{\alpha n'm'\alpha'}+O\left(\frac{1}{N^{2}}\right)\label{eq:GF2D}\end{eqnarray}
 where we have defined the bare $\left(t'=0\right)$ propagator and
the propagator at RPA level respectively as \begin{eqnarray}
G_{\parallel\alpha l'} & = & -\left[\av{\boldsymbol{c}\boldsymbol{c}^{\dagger}}_{\parallel}\right]_{\alpha l'}\\
\tilde{G}_{\alpha l'} & = & -\left[\av{\boldsymbol{c}\boldsymbol{c}^{\dagger}}_{\parallel}^{-1}+\mathbf{t}'\right]_{\alpha l'}^{-1}\label{eq:G_RPA_mat}\end{eqnarray}
 as well as the four point function \begin{eqnarray}
\tilde{\Gamma}_{l'n'm'i'} & = & \mathbf{t}'_{mm'}\left[G_{\parallel l'l}^{-1}G_{\parallel n'n}^{-1}G_{\parallel i'i}^{-1}\av{c_{l}c_{n}c_{m}^{\dagger}c_{i}^{\dagger}}_{\parallel C}\right]\nonumber \\
\label{eq:Ver_RPA_mat}\end{eqnarray}
 where $\av{...}_{\parallel C}$ stands for the connected correlator.

In standard notation with $k=\left(i\omega_{n},\mathbf{k}_{\parallel},\mathbf{k}_{\perp},\sigma\right)$
and $\int_{q}=\frac{1}{\beta}\sum_{\omega_{n}}\sum_{\mathbf{q},\sigma}$,
expression (\ref{eq:GF2D}) translates to \begin{eqnarray}
G\left(k\right) & = & \left[\tilde{G}\left(k\right)^{-1}-\frac{1}{N}\int_{q}\tilde{G}\left(q\right)\tilde{\Gamma}_{4}\left(k,q\right)\right]^{-1}\nonumber \\
 &  & +O\left(\frac{1}{N^{2}}\right),\label{eq:G_next_to_RPA}\end{eqnarray}
 where expressions (\ref{eq:G_RPA_mat}) and (\ref{eq:Ver_RPA_mat})
are respectively given by \begin{eqnarray}
\tilde{G}\left(k\right) & = & \left[G_{\parallel}\left(k_{\parallel}\right)^{-1}-t'\left(\mathbf{k}\right)\right]^{-1},\\
\tilde{\Gamma}_{4}\left(k,q\right) & = & t'\left(\mathbf{q}\right)\Gamma_{1D}\left(k_{\parallel},q_{\parallel}\right),\end{eqnarray}
 with $k_{\parallel}=\left(i\omega_{n},\mathbf{k}_{\parallel},\sigma\right)$
and where $G_{\parallel}\left(k_{\parallel}\right)$ is the one-dimensional
Green's function. The one dimensional quantity \begin{eqnarray}
\Gamma_{1D}\left(k_{\parallel},q_{\parallel}\right) & = & \frac{\av{c_{k_{\parallel}}c_{q_{\parallel}}c_{q_{\parallel}}^{\dagger}c_{k_{\parallel}}^{\dagger}}_{\parallel C}}{G_{\parallel}\left(q_{\parallel}\right)G_{\parallel}\left(k_{\parallel}\right)G_{\parallel}\left(k_{\parallel}\right)}\label{eq:vet_1D}\end{eqnarray}
 is given as a function of the 1D form factors and propagators. Eqs.
(\ref{eq:G_next_to_RPA}-\ref{eq:vet_1D}) permit to obtain the 2D
propagator as a function of the 1D quantities only. This expression
involves higher order correlation functions for the Hubbard chain
which are not known at this point.

\end{document}